\documentclass[superscriptaddress,
tightenlines,
nofootinbib,
12pt,
aps,
prd,
longbibliography,
]{revtex4-1}

\usepackage{graphicx} 
\usepackage[dvipsnames]{xcolor}
\usepackage[colorlinks = true,
            linkcolor = blue,
            urlcolor  = blue,
            citecolor = blue,
            anchorcolor = blue]{hyperref}
\RequirePackage{amssymb}
\usepackage{wrapfig}
\usepackage[separate-uncertainty=true]{siunitx} 
\usepackage{siunitx}

\newcommand{\UZH}{Physik-Institut, University of Zurich, Switzerland}
\newcommand{\LLNL}{Lawrence Livermore National Laboratory, Livermore, CA USA}
\newcommand{\VTNSI}{Virginia Tech National Security Institute, Virginia Tech, Blacksburg VA USA}
\newcommand{\VTPHYS}{Physics Department, Virginia Tech, Blacksburg VA	USA}
\newcommand{\CNP}{Center for Neutrino Physics, Virginia Tech, Blacksburg, VA	USA}

\begin{document}

\title{Nuclear recoil detection with color centers in bulk lithium fluoride}

\collaboration{PALEOCCENE collaboration}

\author{Gabriela R.	Araujo}
\thanks{lead author}
\affiliation{\UZH}
\author{Laura	Baudis}
\affiliation{\UZH}
\author{Nathaniel	Bowden}
\affiliation{\LLNL}
\author{Jordan	Chapman}	
\affiliation{\VTNSI}
\author{Anna	Erickson}
\affiliation{George W. Woodruff School of Mechanical Engineering,
Georgia Institute of Technology, Atlanta, GA USA}
\author{Mariano	Guerrero Perez}
\affiliation{\VTPHYS} 
\author{Adam A.	Hecht}
\affiliation{University of New Mexico, Albuquerque, NM	USA}
\author{Samuel C. Hedges}
\affiliation{\CNP}
\author{Patrick	Huber}
\email{pahuber@vt.edu}
\affiliation{\CNP}
\author{Vsevolod	Ivanov}
\affiliation{\VTNSI}
\affiliation{\VTPHYS}
\affiliation{Virginia Tech Center for Quantum Information Science and Engineering, Blacksburg, VA USA}
\author{Igor	Jovanovic}
\affiliation{Department of Nuclear Engineering and Radiological Sciences, University of Michigan, Ann Arbor, MI	USA}
\author{Giti A.	Khodaparast}
\affiliation{\VTPHYS}
\author{Brenden A.	Magill}
\affiliation{\VTPHYS}
\author{Jose Maria	Mateos}
\affiliation{Center for Microscopy and Image Analysis, University of Zurich,	Switzerland}
\author{Maverick Morrison}
\affiliation{\VTPHYS}
\author{Nicholas W. G.	Smith}
\affiliation{\VTPHYS}
\author{Patrick Stengel}
\affiliation{Jo\v{z}ef Stefan Institute,  Ljubljana, Slovenia}
\author{Stuti	Surani}
\affiliation{Pennsylvania State University, University Park, PA	USA}
\author{Nikita	Vladimirov}
\affiliation{URPP Adaptive Brain Circuits in Development and Learning (AdaBD), University of Zurich, Switzerland}
\author{Keegan Walkup}
\affiliation{\CNP}
\author{Christian Wittweg}
\affiliation{\UZH}
\author{Xianyi	Zhang}
\affiliation{\LLNL}

\date{\today} 

\begin{abstract}

We present initial results on nuclear recoil detection based on the fluorescence of color centers created by nuclear recoils in lithium fluoride. We use gamma rays, fast and thermal neutrons, and study the difference in responses they induce, showing that this type of detector is rather insensitive to gamma rays.  We use light-sheet fluorescence microscopy to image nuclear recoil tracks from fast and thermal neutron interactions deep inside a cubic-centimeter sized crystal and demonstrate automated feature extraction in three dimensions using machine learning tools. The number, size, and topology of the events agree with expectations based on simulations with TRIM. These results constitute the first step towards 10-1000\,g scale detectors with single-event sensitivity for applications such as the detection of dark matter particles,  reactor neutrinos, and neutrons.
\end{abstract}

\maketitle

\section{Introduction}
\label{sec:intro}

The detection of neutral particles with non-zero mass, such as neutrons, neutrinos, and dark matter, usually relies on sensing the signatures of nuclear recoils resulting from particle interactions. The majority of detectors use prompt scintillation, ionization, or phonon signals (or any combination thereof). Instead, we investigate the formation of permanent changes to the crystal structure of the detector material. Specifically, we are interested in the formation of color centers, since these can be detected via fluorescence techniques. Single color centers have been successfully observed, for instance, in diamond~\cite{NVconfocal} using confocal microscopy. While providing the necessary sensitivity, confocal microscopy is constrained to very small volumes measured in cubic micrometers. It has been noted~\cite{Cogswell:2021qlq} that achieving single color center detection capability in macroscopic volumes of cubic centimeters would result in significant sensitivity for rare event searches, {\it e.g.} dark matter or neutrinos. For these applications, it is also important to demonstrate that color center formation purely due to electromagnetic radiation, especially gamma rays, is strongly suppressed in order to achieve the required very low background levels.

Commercial products for radiation dosimetry exist that use the detection of fluorescence via confocal microscopy~\cite{Akselrod:2006a}. These detectors, like all other currently used fluorescent nuclear track detectors (FNTD), {\it e.g.} Refs.~\cite{Akselrod:2011,Akselrod:2018,Akselrod:2020}, are based on charge transfer reactions whereby ionization produces free charge carriers, which then get absorbed by dopants acting as charge traps. These trapped charges are the source of fluorescence; only the electronic configuration of the crystal has changed, but not the underlying crystal structure as defined by the ion lattice. 
These types of materials are entirely unsuitable for rare event detection due to their susceptibility to ordinary ionizing radiation.

Lithium fluoride (LiF) has been studied as a nuclear fluorescent track detector where the fluorescence is directly caused by lattice defects and not by charge transfer reactions. High-quality data on tracks induced by alpha particles, protons, heavy ions, thermal neutrons, and fast neutrons have been reported by the Bilski group~\cite{Bilski:2017,Bilski:2018,Bilksi:2019a,Bilski:2019b,Bilski:2024ghu}. These results have been obtained using traditional wide-field fluorescent microscopy, which limits the volume of samples that can be measured.  What makes LiF a promising target is that the excitation wavelength of 450\,nm and the emission wavelengths of 525\,nm and 650\,nm are well separated and are in the visible range. Together with the considerable emission brightness observed by the Bilski group, these features make LiF an ideal material for our research.

Moreover, first-principles calculations have established a clear theoretical picture of the underlying changes to the crystal structure~\cite{Perez:2024hly}: the 525\,nm emission corresponds to a positively charged complex of three fluorine vacancies ($F^+_3$) whereas the 650\,nm emission is due to a neutral two fluorine vacancy complex ($F^0_2$). Our data supports these results, and we find clear differences in the rate of formation of these two types of color centers based on the different responses to neutron and gamma irradiation. We further find that for a given dose, the fluorescence response to gamma rays from a $^{60}$Co source is about 50 times less than to fast neutrons from an AmBe source.

Furthermore, we demonstrate the imaging of paired tracks created by an alpha particle and triton which result from thermal neutron capture on $^{6}$Li as well as  fast neutron recoils in three dimensions using light-sheet microscopy several millimeters deep inside the sample. We used a mesoSPIM~\cite{Vladimirov:2024} for this demonstration; this microscope can scan large samples, tens of cubic centimeters, and reach a scan speed of less than ten minutes per cubic centimeter of sample for an isotropic resolution of $\sim$\SI{4}{\um}. This unlocks the ability to scan large volumes to reach single event detection sensitivity in detectors of masses in the range of 10-1000\,g.  We also show cosmic ray event candidates, which, given their size, could be the result of muon or very high-energy neutron interactions. This constitutes the first step toward the use of LiF as a direct dark matter detector, sensitive neutron detector, and reactor neutrino detector~\cite{Cogswell:2021qlq}.

\section{Radiation damage}
\label{sec:radiation}

The neutron source we use is an americium-beryllium (AmBe) source with an activity of 3.7\,GBq producing  approximately 220\,000 neutrons per second, and the gamma source is a $^{60}$Co source with an activity of 2.8\,GBq; both sources were manufactured by QSA Global.  We use a GEANT4~\cite{GEANT4} simulation based on the actual sample-source geometry to determine the dose transferred to our samples: we find a mean dose of 6.6\,pGy per neutron\footnote{The neutron source is enclosed in a 7.5\,mm lead shield to absorb all x-rays produced in the decays of $^{241}$Am.} and a mean dose of 9\,pGy for the two gammas from the $^{60}$Co source, {\it i.e.}, the dose rate from the gamma source is about $17\,000$ times higher than from the neutron source.

\begin{figure}[tb]
    \centering
    \includegraphics[width=0.7\linewidth]{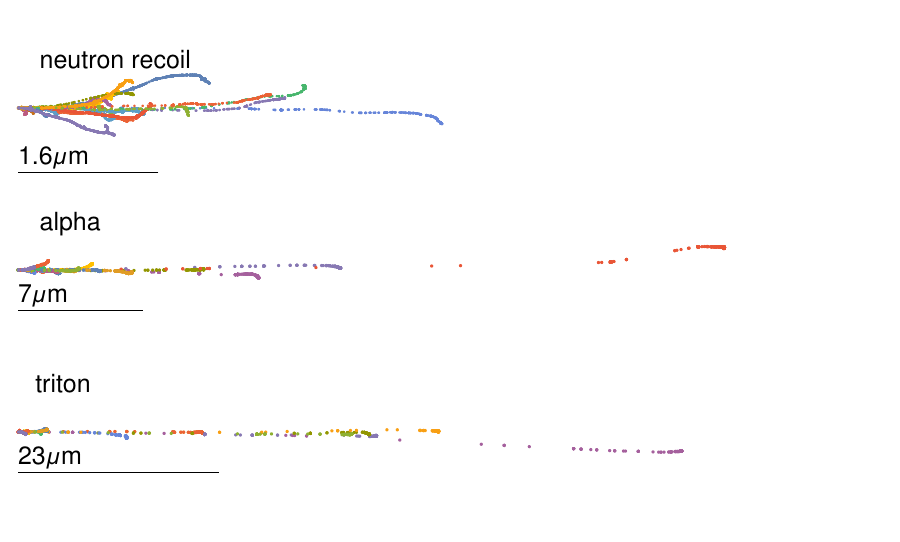}
    \caption{Simulation of vacancies created in LiF using TRIM. Shown are 20 events each for fast-neutron induced recoils, alpha particles, and tritons produced by thermal neutron capture on $^6$Li. Note the different scales for each particle type. Bars indicate scale and mean range.}
    \label{fig:trim}
\end{figure}

 To estimate the track sizes and topology we perform a simulation of vacancy production in LiF using TRIM~\cite{TRIM} for three types of events: fast neutron-induced recoils where the neutrons follow the spectrum of an AmBe source~\cite{ambe}, alpha particles of 2.05\, MeV energy and tritons of 2.73\, MeV energy as produced in thermal neutron capture on $^6$Li. The capture of thermal neutrons on $^6$Li, with a natural abundance of 7.6\%,  produces pairs of back-to-back and mono-energetic alphas and tritons via the reaction
\begin{equation}
\label{eq:fission}
    {}^6\textrm{Li}+n\to\alpha~(2.05~\textrm{MeV)} + {}^3\textrm{H}~(2.73~\textrm{MeV)}\,,
\end{equation}
with a cross section of 940\,b~\cite{Sears01011992}. These events occur throughout the entire volume and the directions are isotropic.

 The AmBe neutron source may also act as a thermal neutron source by introducing a suitable moderator, in our case high-density polyethylene (HDPE). Based on a GEANT4 simulation of the neutron source, its borated HDPE shield, and the HDPE moderator we estimate a flux of thermal neutrons of about $160-170\,\mathrm{s}^{-1}\,\mathrm{cm}^{-2}$ compared to a flux of fast neutrons of approximately $1100\,\mathrm{s}^{-1}\,\mathrm{cm}^{-2}$ at the sample position.

 Figure~\ref{fig:trim} shows examples of vacancy trails produced by these interactions together with their mean range. For combined alpha-triton tracks we find a mean range\footnote{The mean range is not simply the sum of the mean range of each event type since the underlying distributions are not symmetric and have  long tails.} of 32\,$\mu$m. The approximate mean density of fluorine vacancies formed along the track for a neutron-induced recoil event is 1200$\,\mu\mathrm{m}^{-1}$, for an alpha event it is 100$\,\mu\mathrm{m}^{-1}$ and for a triton event it is 30$\,\mu\mathrm{m}^{-1}$. The primary optically active color centers in LiF are the two and three vacancy complexes, $F_2^0$ and $F_3^+$; thus, the number of color centers is 1/3 to 1/2 of those numbers. Remarkably, for triton events, this translates to a mean density as low as 10-15$\,\mu\mathrm{m}^{-1}$ color centers. For the track itself, without the Bragg peak, the mean density becomes 3-5$\,\mu\mathrm{m}^{-1}$ color centers. However, as we will show, the Bragg peak and track are clearly visible despite the relatively high noise level of the camera. The Bragg peak for alpha particles is predicted to be 2.5 times brighter than for tritons, whereas neutron-induced recoil should yield events about 6 times brighter than those from alpha particles.

Most LiF samples were exposed to 1-2 years of cosmic ray neutrons at 634\,m elevation (the elevation of Blacksburg, VA) and about 8 hours of transatlantic flight at about 10\,000\,m altitude.  The elastic neutron scattering cross sections for fast neutrons are 1.4\,b (Li) and 4.018\,b (F)~\cite{Sears01011992}. The density of LiF is $2.64\,\mathrm{g}\,\mathrm{cm}^{-3}$ and the molar mass is 25.95\,u, which corresponds to $6.13\times10^{22}$ Li and F atoms per cubic centimeter, each. For the cosmic ray neutron flux estimate we follow Ref.~\cite{Ziegler:1996}: the flux of fast neutrons falls as $E^{-1.5}$ and for the range from 10-1000\,MeV the flux is $1.75\times10^5\,\mathrm{cm}^{-2}\,\mathrm{yr}^{-1}$, thus for lower cutoff energies this number increases correspondingly. The same reference also provides simple scaling laws with altitude; for 634\,m altitude, the increase is a factor of 1.6, and for 10\,000\,m the increase is a factor of 122. The 8 hours of transatlantic flight correspond to about 1 month at 634\,m. Overall, the prediction for fast neutron events from cosmic rays ranges from $500-3\,300$ events per $1\,\mathrm{mm}^3$ of LiF depending on the lowest cutoff neutron energy capable of creating a visible event ranging from $10-0.1\,$MeV, respectively. 

\section{Bulk spectroscopy}
\label{sec:spectrum}

\begin{figure}[t]
    \centering
  
    \includegraphics[width=0.6\textwidth]{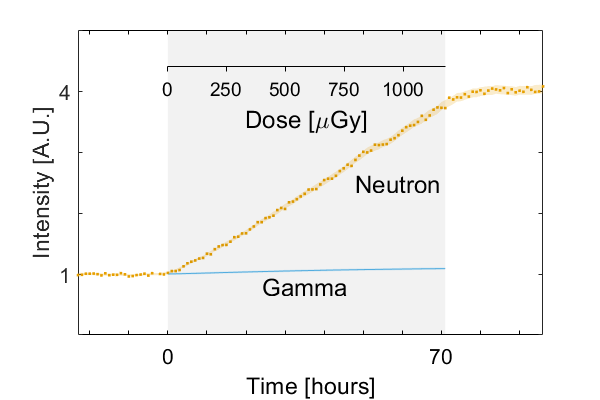}   
    \caption{LiF fluorescence signal before, during and after neutron irradiation. The highlighted area indicates the period of the irradiation.}
    \label{fig:fiber}
\end{figure}

In order to test the difference in response of LiF to neutron and gamma irradiation, we performed a series of irradiation experiments: for all experiments, we use polished 1\,cm$^3$ LiF cubes from two different suppliers (United Crystal and Crystran). The initial, {\it i.e.} prior to any irradiation, fluorescence strength under excitation at 450\,nm varies considerably between batches, but we do not observe a significant difference in radiation response between samples from the two suppliers or between batches.

In the first series of measurements, we use optical fibers to deliver the excitation light and to collect the fluorescence signal to be able to perform the fluorescence measurement concurrently with irradiation. The samples are kept under a dry nitrogen cover during the entire experiment to prevent humidity from affecting them. The excitation light is provided by a 450\,nm laser diode (Thorlabs NPL45B) and coupled into a fiber (Thorlabs M124L02). We couple about 3\% of the laser light directly into a fiber after a series of neutral density filters (Thorlabs NDUV10A/NDUV40A) and then onto one of our SiPM detectors to monitor the output stability of the laser diode. At the end of the fiber delivering the laser to the sample, a $450\pm10$\,nm bandpass filter (Thorlabs FBH450-10) is used to remove any fluorescence occurring in the fiber.  The collection fiber has an input low-pass filter (Thorlabs FESH0500 + FBH600-40) to prevent scattered laser light from entering the fiber. We use a collimator at the end of each fiber to provide a controlled optical geometry. The control channel and collection fibers are each directed onto their own SiPM (Hamamatsu MPPC S13360-1350CS) operated at $-19^\circ$C. The full waveforms from the SiPMs are digitized using a 14-bit two-channel 62.5\,MHz CAEN digitizer. We use a trigger generator to pulse the laser and trigger data acquisition at a rate of 20\,kHz and integrate photon counting data for 200 seconds in 5 laser-on periods interleaved with 5 laser-off periods starting every hour. We then perform an on-off background subtraction to eliminate the SiPM noise floor. The resulting data is shown in Fig.~\ref{fig:fiber}. In orange, we show the data for neutron irradiation as a function of time, with the shaded region indicating the period when the radiation source was present. The cyan line is for the same type of data but for gamma irradiation scaled to the same dose rate as the neutron irradiation. The two slopes are different by a factor of 50, indicating that these detectors are relatively immune to electromagnetic ionizing radiation. It should be noted that the increase in fluorescence signal continues for a few hours after the end of the neutron irradiation, but this is a relatively small increase.

\begin{figure}[tb]
\includegraphics[width=0.48\textwidth]{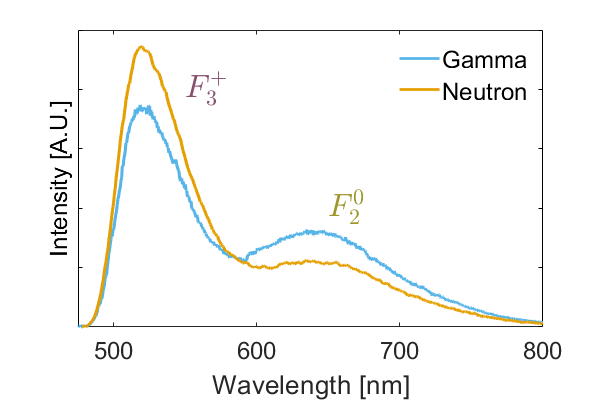}
\includegraphics[width=0.48\textwidth]{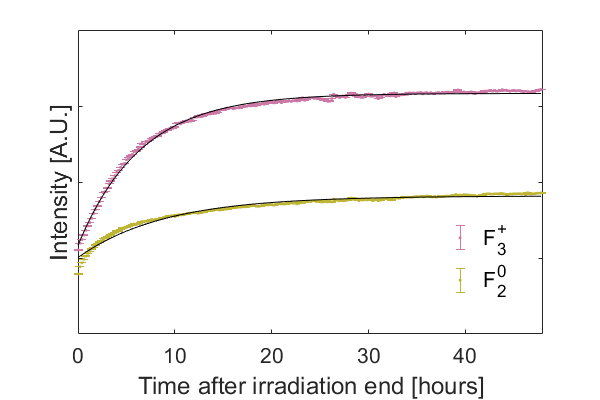}    
\caption{Left-hand panel: the fluorescence from 450\,nm excitation of LiF after exposure to gamma and neutron irradiation normalized to the same area. Right-hand panel: this LiF sample was exposed to 1 hour of gamma radiation and then observed at room temperature in 12-minute intervals after gamma exposure. The fluorescence signals are integrated over the spectral range covering the $F_2^0$ and $F_3^+$ regions.}
\label{fig:spec}
\end{figure}

To further investigate the difference between neutron and gamma response, we study the spectral response and relaxation times. These data were obtained using an Agilent Cary Eclipse UV-VIS G9800A fluorospectrometer with an excitation wavelength set to $450\pm20$\,nm. We also use a 500\,nm cut-on long pass filter (Thorlabs FELH0500) to prevent any scattered excitation light from interfering with the measurement. In Fig.~\ref{fig:spec}a, we show the fluorescence spectrum after either irradiation normalized to the same area. We clearly can see that gamma radiation leads to more formation of $F^0_2$ centers compared to neutron radiation.

In Fig.~\ref{fig:spec}b we show the time evolution of the integrated fluorescence signal after a 1-hour gamma irradiation and observe an exponential increase in signal strength with time constants of $6.5$\,h for the $F^3_+$ center and a time constant of $9$\,h for the $F^0_2$ center. The spectral range for integration for the $F^3_+$ center is $485-585$\,nm and for the $F^0_2$ center, it is $585-685$\,nm. For neutrons, this relaxation effect is much less pronounced, as can be seen in Fig~\ref{fig:fiber}, and happens faster.

One plausible interpretation of these results is that gamma rays have a small probability of creating a single fluorine vacancy ($F$ center). We do not detect these $F$ centers optically despite several separate measurements using the wavelengths computed in Ref.~\cite{Perez:2024hly}. $F$ centers can diffuse through the crystal to form optically active $F^0_2$ and $F^3_+$ centers. For neutron irradiation, a large number of $F$ centers are formed locally; hence, the distance over which they need to move is rather short, and therefore, the full fluorescence strength is formed fairly promptly.

We find that the fluorescence signal stays stable for periods exceeding several months for samples stored under a vacuum at room temperature. The LiF crystal is slightly hygroscopic, and the vacuum (or dry nitrogen) prevents humidity from interacting with the samples. Without this precaution, a slow decline of the fluorescence signal over a period of days is observed that eventually plateaus. Furthermore, we performed studies of temperature response by heating samples for several hours {\it in vacuo} at a range of temperatures from room temperature up to $350^\circ$C. We find that the post-irradiation signal remains stable for temperatures up to $100^\circ$C. Continued heating to $350^\circ$ C resets the fluorescence to its pre-irradiation levels, {\it i.e.} the radiation damage is annealed.

\section{Light-sheet fluorescence microscopy}
\label{sec:spim}

\begin{figure}[tb]
    \centering
    \includegraphics[width=0.75\textwidth]{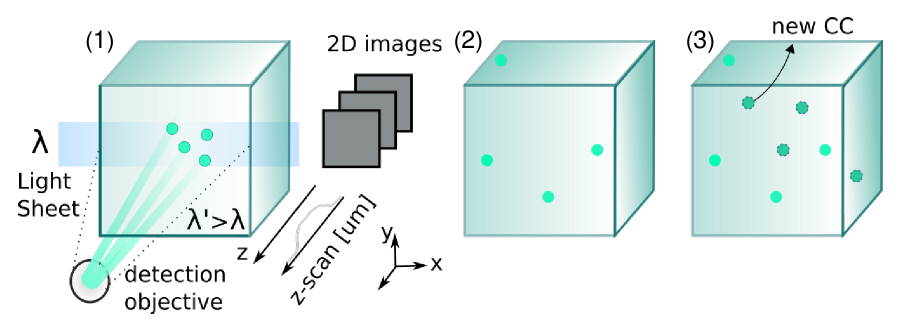}
    \caption{In LSFM, a thin light sheet illuminates a z-section of the sample x-y-plane, which is observed by a camera orthogonal to the light-sheet source (1). Filters are employed before the camera to allow only fluorescence light to reach the camera. By incrementally shifting the sample along the z-axis, $3$D images are generated. To understand signals originating from the irradiation procedure, crystals are measured before (2) and after exposure (3).}
    \label{schematic_lsfm}
\end{figure}

In fluorescence microscopy, a sample is excited with a specific wavelength of light and the emitted fluorescent light is collected using a combination of filters, objective lenses, and a camera. By filtering out the excitation wavelength, the signal emitted by fluorescent color centers can be effectively isolated in the images. The brightness of each imaged pixel can provide information about the number of color centers within that pixel, possibly allowing for the differentiation between single-site vacancies and full tracks.
We use light-sheet fluorescence microscopy (LSFM, also known as SPIM, selective plane illumination microscopy), shown in Fig.~\ref{schematic_lsfm}, due to its high volume throughput.

Although single color centers have been imaged using wide-field and confocal fluorescence microscopes \cite{doi:10.1080/23746149.2020.1858721, D0NR05931E}, imaging color centers with LSFM offers further advantages: optical sectioning of bulk three dimensional samples compared to wide-field microscopy, reduced sample exposure (minimizing bleaching of signals), higher throughput and working distance \citep{Vladimirov:2024}. These features make LSFM a well-suited readout technology for color centers in large crystals. 

For the measurements described in this work, we use a benchtop microscope from the mesoSPIM initiative ~\citep{webmesospim, Voigt2019, Vladimirov:2024}, an open-source project comprising over 35 microscopes worldwide.  The mesoSPIM microscopes offer near-isotropic resolution imaging of cm-sized samples within minutes, eliminating the need to slice the sample as often required for large sample imaging. This capability is attributed to the long working distances of the objectives and the light-sheet optical sectioning. This allows for imaging color centers in three dimensions, deep within crystals. The mesoSPIM resolution and scan speed can be adjusted using a set of automated lenses (with magnifications ranging from $0.63\times$ to $20\times$) and the light-sheet axial z-step. 

The current version of the mesoSPIM, shown in Fig.~\ref{meso}, attains \SI{1.5}{\um} resolution in the x-y-plane and \SI{3.3}{\um} in the axial direction. The latter is defined by the minimum thickness of the light sheet. The setup can reach a scan speed of less than ten minutes per cubic centimeter of sample for an isotropic resolution of $\sim$\SI{4}{\um}. 

In the mesoSPIM used in these measurements, a \SI{445}{\nm} laser (Cobolt 06-MLD 50 mW, H\"ubner Photonics) provides the excitation light, while an sCMOS camera (Photometrics Iris15) records the fluorescence, with a \SI{488}{\nm} long-pass emission filter (RazorEdge LP Edge Filter 488 RU, AHF Analysentechnik AG, Article-Nr. F76-488). The light sheet is created by an electro-tunable lens (Optotune EL-16-40-TC-VIS-5D-C) and is swept axially. This axial sweeping method, known as axially scanned light-sheet microscopy mode (ASLM), yields uniform z-resolution across the field of view by translating the thinnest part of the light sheet (the beam waist) through the sample in sync with the rolling shutter of the camera. The camera's field of view is 2960$\times$5096 pixels large, with a pixel size of \SI{4.25}{\um}. With 16-bit depth values acquired per pixel, each image yields 30\,MByte of data.

\subsection{Methods}
\begin{figure}[tb]
    \centering
    \includegraphics[width=\textwidth]{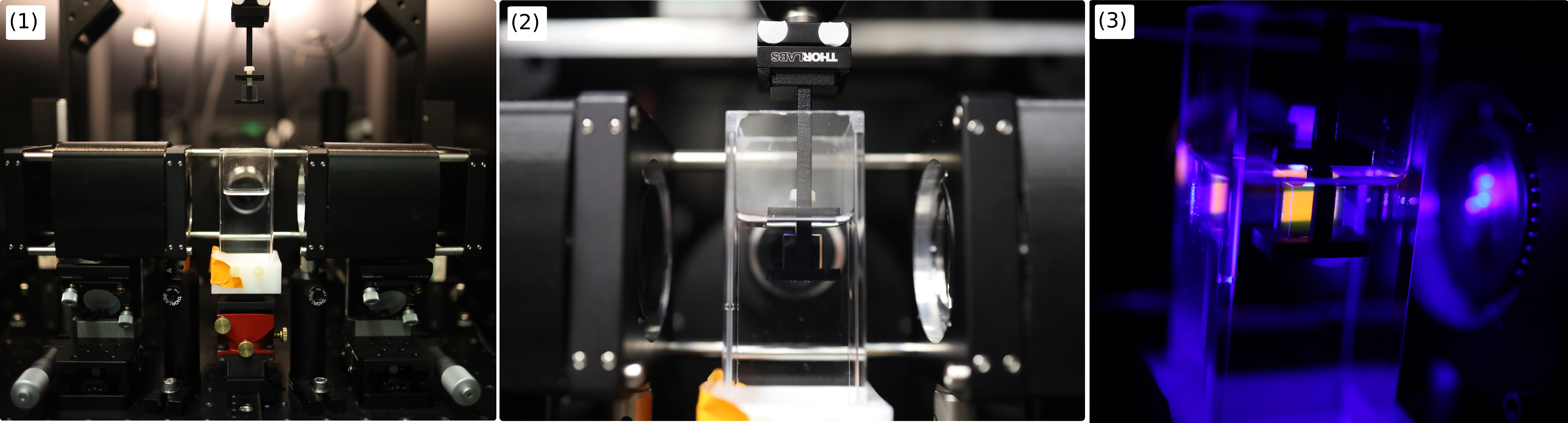}
    \caption{(1) Loading a LiF crystal into the immersion chamber of the mesoSPIM microscope. (2) Crystal inside the immersion medium. (3) When the light sheet is on, its light is observed as scattered light in the medium and fluorescent light inside the crystal (in this case, a highly irradiated crystal).}
    \label{meso}
\end{figure}

The LiF crystals were irradiated with neutrons with intensities as described in Tab.~\ref{tab:samples}.
The LiF crystals were placed in a custom-made sample holder designed for imaging crystals with the mesoSPIM, see Fig.~\ref{meso}(1). To circumvent the need to refocus when moving the crystal in the z-direction, the crystal was immersed in a refractive-index matching medium (Cargille \#1803, nD=1.395). The mesoSPIM Python-based software was used to control the microscope and acquire the images. These included focusing and translating in x, y, z or rotating the sample, selecting excitation wavelength,  long-pass filter, laser intensity, and galvo amplitude. The latter is associated with the coverage area of the light sheet, Fig.~\ref{meso}(3). The imaged area outside the galvo coverage is used to assess the camera background, while the rest, here called \textit{luminous} area, is utilized for the analysis of the signal, and fluorescent tracks.

To precisely calibrate the setup and analysis software for these measurements, nanometer-sized fluorescent beads, and quantum dots immersed in a transparent medium, epoxy or agarose, were imaged utilizing 5x, 10x, and 20$\times$ magnification. These measurements confirmed the stage repeatability and constant focus across z-planes when moving the samples and were used to assess the microscope point spread function for all the studied magnifications. 

When imaging the crystals, it was noted that the color centers from tracks in LiF were much dimmer than the fluorescent beads and quantum dots. The latter were sharply imaged using camera exposure times of only 20\,ms, while the tracks required between one and two orders of magnitude longer exposure to yield a clear signal. Increasing the laser power did not help in increasing the signal, as the tracks bleached upon exposure to higher laser intensities. Given that reducing camera noise was not possible, we then opted for utilizing moderate laser intensity at longer exposure times (up to five seconds per image). Applying such prolonged exposure required moving from the ASLM mode of imaging to single full-image acquisitions -- as repeated images in sync with the shutter would require too much time and yield bleaching of color centers in the tracks.

In no-ASLM mode, sufficient signal from tracks was obtained at acceptable bleaching rates and satisfactory z-resolution (albeit not uniform). At 10$\times$ magnification, most tracks were larger than the full thickness of the z-section, which was $\sim$\SI{4}{\um} at the beam waist at the center and slightly larger at the edges. To exclude the regions of the largest spread of the light sheet ($\sim$\SI{10}{\um}), sections of the data closer to the center are utilized. As a result, the intensity across a full image is non-uniform, preventing any calorimetric analysis of event brightness or energy.

\subsection{Data analysis}

\begin{figure}[htb]
    \centering
    \includegraphics[width=0.38\linewidth]{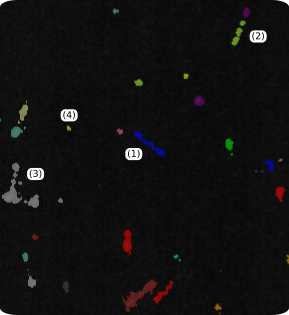}
    \caption{Example of segmentation by Ilastik - Segmented signal candidates, clustering of pixels above the threshold, as predicted by the algorithm using labeled sets, are shown in colors. A blue-labeled track is observed at the center (1), and some not well-segmented tracks are also shown for illustration: Track (2) was segmented into two objects instead of one; track labeled as (3) is probably the result of 2 tracks that merged and track (4) might be just noise or a small clustering of color centers from low-energy interactions. In either case, we exclude these using a feature size threshold that depends on the tracks to be studied. }
    \label{label_example}
\end{figure}

With the scans of several crystals irradiated with different particles and fluxes, over one terabyte of data was acquired. Therefore, a software-automatized analysis was developed. 
The analysis consisted of the following steps:  
\begin{enumerate}
    \item  Raw data selection and pre-processing: In this step, the \textit{luminous} data of each scan is separated from background data (as mentioned in the previous section), and the edges of the image are discarded. A Gaussian blurring filter is applied to the final dataset to enhance the signal within the noise.
    \item Feature segmentation \& classification: We use the open-source software Ilastik~\cite{berg2019} to segment the image, effectively separating track candidates from the background, as shown in Fig.~\ref{label_example}. This process is performed by labeling a few tracks and applying the prediction of the algorithm to the rest of the data set as well as other datasets.
    \item Analysis (part I): In this step, we apply the segmentation mask to the data and calculate feature properties using the Python skimage package~\cite{scikit-image}. In the masking procedure, the background is masked with zeros and tracks above a feature size threshold to keep their intensity values. The feature size threshold corresponds to the minimum number of connected pixels in the cluster (as calculated by the `remove small objects' function from skimage.morphology), and it varies depending on the analysis type: for the analysis of large tracks (such as those induced by thermal neutrons), the threshold is tuned to suppress small features (those with a major axis length smaller than $\sim$\SI{4}{\um}, as measured by skimage.measure). For the analysis of smaller tracks (such as those caused by fast neutrons), the threshold is greatly loosened (to $\sim$ 5\% of the previous value), such that only small noise clusters are cut (\textit{low-size threshold} data).   
    \item Post-processing \&  quantitative analysis (part II):
    As the software-predicted segmentation often yields wrong segmentation and/or classification, we conduct the correction of feature classification in a post-processing step. This is necessary because the signal-to-noise ratio is not high, the data acquisition parameters and resolution are not fully uniform, and the software has not yet been fully optimized. This final selection of tracks is made manually by removing (or classifying into a new category) merged tracks, such as item (3) in Fig.~\ref{label_example}, tracks that are not straight, tracks split into multiple parts, {\it e.g.} item (2) in Fig.~\ref{label_example} and tracks that appear too fuzzy or look like noise. After this final classification, the intensity distribution along the track is analyzed, as well as the track direction and size. To estimate the track size, we neglect the fact that the axial resolution is not homogeneous; therefore, the track sizes reported here are good estimators only for tracks that are mostly perpendicular to the z-axis.
\end{enumerate}

Data for which only step (1) was applied, is named throughout as \textit{pre-processed}, while \textit{processed} data is referred to as `masked' data, that is, after the processing described in (3). As the data is a three-dimensional array of intensities, the visualization of large volumes in this form is difficult. Therefore, we will often present the data in two-dimensional \textit{z-stacks}.  In a \textit{z-stack}, the intensities from all the z-planes are projected into a single plane. For \textit{processed} data, only the intensities within the non-zero mask are summed. Two-dimensional z-stacks are only used to display data, segmentation and quantitative analysis, {\it e.g.} size determination, are performed on the underlying three-dimensional data.

We note that the acquired images correspond to large areas ($\sim$\SI{2}{\mm\squared}) spanning several z-planes with axial depths up to 1\,mm, including thousands of tracks. However, to facilitate visualization, we show only small sections of a few datasets.  As mentioned in the previous section, the observation of luminous tracks is only expected within the galvo illuminated region, which corresponds to roughly 40\% of the camera chip size, with specific values depending on the specific data set. To estimate the current scan speed achieved for the results presented here, we note that the exposure time per z-plane was 5\,s. The volume of one z-plane scan is given by the size of the z-step, typically $6\,\mu$m, multiplied by the area of a pixel of $4.25\times4.25\,\mu\mathrm{m}^2 $, the number of pixels 2960$\times$5096 ($\times$ 40\%) and divided by the square of the magnification, which in our case $10\times$, yielding a total of $6.5\times10^6\,\mu\mathrm{m}^3$ per z-plane. This corresponds to a scan time of about 200\,h for a cubic centimeter. With a magnification of $10\times$ the lateral resolution is about $1.5\,\mu$m, three times larger than the pixel size (0.425\,$\mu$m). Reducing the magnification by a factor of three would reduce the scan time by a factor of nine without losing much information (while also increasing the signal-to-noise ratio per pixel). A lower noise camera, {\it e.g.} a Hamamatsu Orca Quest 2.0 would provide a much lower noise level and thus reach the same signal-to-noise ratio with the exposure time per frame of about 1/10 of the current value or about 0.5\,s, further allowing for full coverage of galvo exposure (instead of the current 40\%). These improvements point towards plausible scan times of 1-2 hours for $1\,\mathrm{cm}^3$. The associated data volume will be of the order of $1\,\mathrm{TByte}\,/\mathrm{cm}^3$. We should also remark that if materials can be found that are not subject to bleaching, another order of magnitude can be gained by increasing the laser power to 100-150\,mW.

\subsection{Results}

The results in the following are derived from four samples from the same supplier (Crystran) and batch, all samples are cubes $10\times10\times10\,\mathrm{mm}^3$ with 6 optically polished sides. Their histories are summarized in Tab.~\ref{tab:samples}.

\begin{table}[h]
    \centering
    \begin{tabular}{p{4cm}|p{2cm}|p{2cm}|p{2cm}|p{2cm}}
    sample &  0504  & 0505 & 0506 & 0507  \\
    \hline
     radiation & fast neutrons & thermal neutrons & none & none \\
     fluence [$\mathrm{cm}^{-2}$] & $1.7\times10^6$ & $1.3\times10^5$ & none & none\\
     heat treatment & none    & none & none & 6 hours at $350^\circ$C\\
     
    \end{tabular}
    \caption{Samples used for imaging studies. All samples spent about 8 hours on transatlantic flights and were exposed for 1-2 years to cosmic ray neutrons (see text for details). The heat treatment of sample 0507 took place within the week prior of it being scanned and hence it effectively had a negligible cosmic ray exposure. The thermal and fast neutron fluences are approximately matched to induce the same number of events.}
    \label{tab:samples}
\end{table}

\begin{figure}[p]
\centering

    \includegraphics[width=\textwidth]{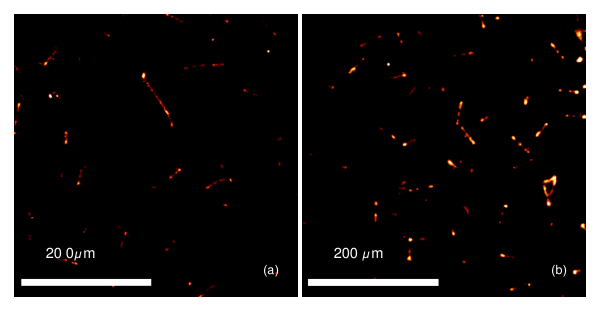}
 \caption{\textit{Processed z-stack} (\SI{0.2}{\mm\squared} x \SI{0.3}{\mm} axial depth) from a non-irradiated crystal (sample 0506) imaged at 10$\times$ magnification (a).
 \textit{Processed z-stack} (0.2 mm$^2$ x 0.1 mm axial depth) from a crystal irradiated with $1.3\times10^5\,\mathrm{cm}^{-2}$  thermal neutrons (b, sample 0505) imaged at 10$\times$ magnification. All images are taken $\sim$\SI{2}{\mm} deep in the crystal. The stacks were processed in the same manner (\textit{low-size threshold} processing), and the data was acquired at similar illumination conditions. }, 
  \label{Thermal_vs_fast_neutrons}
 \end{figure}

The relation between irradiation dose and absolute luminescence of the crystal was established with spectroscopy in Sec.~\ref{sec:spectrum}. Therefore, we present here only a qualitative analysis of the track images corresponding to the different irradiation doses. We start with the results from a crystal that has not been irradiated but has been exposed to cosmic rays for 12-24 months, including during transcontinental airplane flights. Many tracks are observed in panel (a) of Fig.~\ref{Thermal_vs_fast_neutrons}. Some smaller tracks are possibly from neutron-induced nuclear recoils or from $^6$Li neutron capture (see Eq.~\ref{eq:fission}), while the other larger ones were probably induced by other cosmic rays. As a comparison, sample 0507 (not shown), a crystal that was annealed at $350^\circ$C before imaging, showed no significant amount of tracks: no discernible features above the noise level within a volume eight times larger than the one scanned for sample 0506. In panel (b) of Fig.~\ref{Thermal_vs_fast_neutrons} we show the result of irradiation with thermal neutrons of sample 0505. This sample shows a significant number of double-peak extended tracks as expected for the result of thermal neutron capture on $^6$Li. The details of the track shapes and intensity profile of each event type are discussed in the following.

\begin{figure}[p]
\centering
		\includegraphics[height=.4\textwidth]{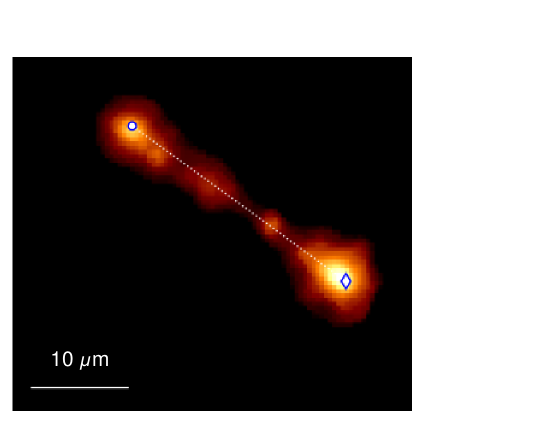}\hspace*{-1.5cm}\includegraphics[height=0.35\textwidth, viewport = 0.00     0.00   260.00   182.0, clip = true]{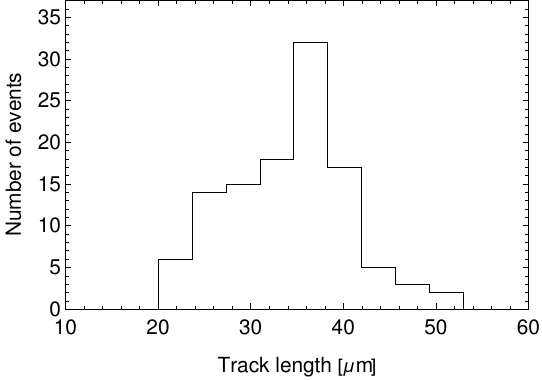}
 \caption{\textit{Processed z-stack} of a thermal-neutron induced track candidate (left) in sample 0505 together with the axis defining the track length. On the right, a histogram of the three-dimensional track length measured for a selection of events from several scans of sample 0505 that meet the quality criteria as defined in the text.
  \label{T_track}}
 \end{figure}

 \begin{figure}[htp]
\centering
        \includegraphics[width=\textwidth]{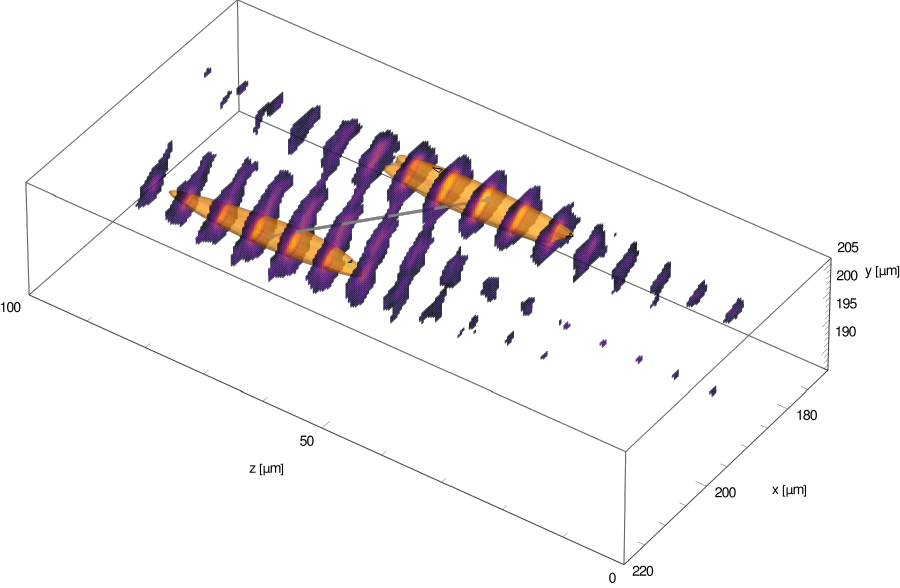}\llap{\raisebox{8cm}{
        \includegraphics[height=5cm]{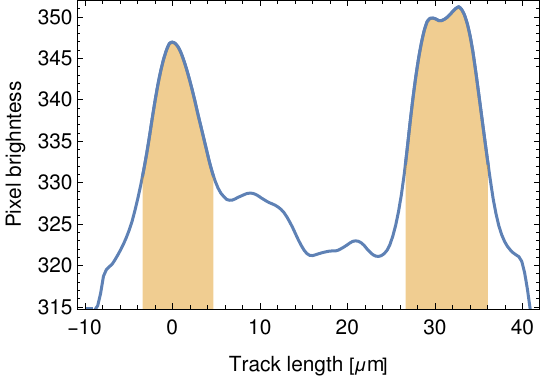}}}
  \caption{Three-dimensional view of a thermal neutron capture events in sample 0505 from pre-processed data. The conversion from pixel/plane coordinates is  0.425\,$\mu$m for the x- and y-axes, and $6\,\mu$m for the z-axis. The color scales with the brightness of the pixel, with darker colors indicating lower brightness. The enclosed contours represent the full-width at half maximum volume. The line connects the points of maximum brightness of each volume.   The inset shows the projection of maximum brightness along this line. The colored/shaded regions correspond to the enclosed contours.  The coordinate origin of the shown volume is the top left-hand corner of the scan volume of $1.13\times1.13\times0.216\,\mathrm{mm}^3$. The scan volume is 2\,mm deep inside a $10\times10\times10\,\,\mathrm{mm}^3$ LiF cube.
  \label{T_track_3D}}
 \end{figure}

As the back-to-back tracks induced by the capture of thermal neutrons by $^6$Li originate at the same point, they should appear as a single track. However, the brightness increases with the distance from the interaction point due to the usual Bethe-Bloch energy losses; in particular, there will be two Bragg peaks: a brighter one for the alpha (higher $dE/dx$) and a somewhat less bright one for the triton. The specific energy loss is lower at the beginning of each track, and the signal level is very close to the noise level; thus, some of these tracks might appear slightly separated. Especially, the triton track is predicted to have a low density of color centers 3-5$\,\mu\mathrm{m}^{-1}$, see Sec.~\ref{sec:radiation}. In most of the features classified as tracks by the automated reconstruction, the highest intensity is observed either at one end of the track length or at both ends, as shown in the example in the left-hand panel of Fig.~\ref{T_track}.

To estimate the length of tracks induced by thermal neutrons with higher accuracy, we select tracks that: i) present the highest pixel brightness intensities at the edges; ii) do not have a large inclination angle with relation to the x-y plane (in this way, we can neglect the correction for the z-inhomogeneity). The average track size obtained from over 100 tracks fulfilling these requirements is \SI{34.2}{\um}, compared to the prediction of $32\,\mu$m. The frequency distribution of track lengths is shown in the right-hand panel of  Fig.~\ref{T_track}

In Fig.~\ref{T_track_3D} we show the three-dimensional distribution of brightness for one selected neutron capture event candidate. We show the brightness distribution in each z-plane.    We also show the iso-contours of one half of the maximum brightness for each of the two maxima. 
There are two well-separated volumes, which likely correspond to the Bragg peaks of the alpha and triton, respectively. The total amount of light above the background contained in the left volume, likely corresponding to the triton Bragg peak, is overall $2.3\times$ less than the for the right volume, presumably the alpha Bragg peak; simulation with TRIM (see Sec.~\ref{sec:radiation}) predicts a factor of 2.5. The line shown connects the two brightest points and we can project the event onto this line as shown in the inset. This distribution closely resembles a $dE/dx$-curve and with more calibration could be used for event identification.

\begin{figure}[tb]
    \centering
    \includegraphics[width=\linewidth]{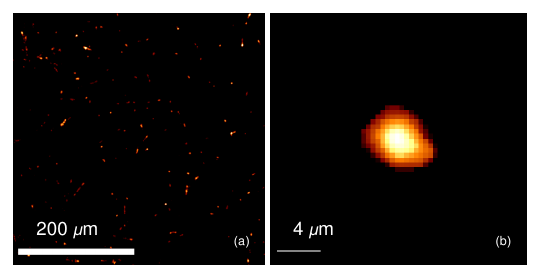}
    \caption{\textit{Processed z-stack} (0.2 mm$^2$ x 0.1 mm axial depth) from a crystal irradiated with $1.7\times10^6\,\mathrm{cm}^{-2}$ fast neutrons (sample 0504) at 10$\times$ magnification on the left. On the right a zoom into a single fast neutron event from the same scan.}
    \label{fig:fast}
\end{figure}

In Fig.~\ref{fig:fast} we show the result of a scan of sample 0504 which was irradiated with fast neutrons. Most events are small elongated clusters of bright pixels about $2-4\,\mu$m in size as expected from simulation. The left hand panel shows a processed z-stack of a larger volume with many events, whereas the right hand panel shows a typical fast neutron event candidate close-up.

Now that we have seen that event morphologies for thermal neutron capture and fast neutron recoil match simulation, we would like to compare the number of thermal neutron capture events as well as of fast neutron recoils from cosmic rays to prediction. We use sample 0505 and we hand-scanned all 943 objects identified by the automated reconstruction within a scanned volume of $1.13\times1.13\times0.216\,\mathrm{mm}^3$. The central third of the volume has been bleached by the waist of the light sheet, and thus no tracks are recorded there; the active volume therefore is approximately $0.18\,\mathrm{mm}^3$.  We sorted the events into three categories: compact single peak,  422 events (c), extended double peak, 136 events, (e) and amorphous, 385 events (a). The latter category contains all events with multiple peaks, tracks with kinks, and several disconnected tracks.   All tracks found in this analysis are shown in the appendix. Together with the thermal neutron flux (see Sec.~\ref{sec:radiation}) and given volume, we expect approximately 100-160 thermal neutron captures compared with 136 events in category (e). We should note that this prediction has large uncertainties, and a precise calibration of the thermal neutron flux is the subject of future work.

Based on the discussion in Sec.~\ref{sec:radiation} we expect 100-600 fast neutron events from both cosmic ray exposure and neutron source exposure compared to 422 observed events in category (c). This very rough agreement clearly requires refinement in future work. The large number of amorphous events in category (a) could be in part due to the mis-identification of events because of limited resolution, non-uniform z-resolution, and bleaching of parts of the event. It seems unlikely that neutron capture reactions on $^7$Li are the cause since the cross sections for these are low~\cite{PhysRev.114.1037,BROWN1963137}. Again, this is an area for future investigation.

\begin{figure}[p]
\centering
	\centering
    \
		\includegraphics[width=0.9\textwidth]{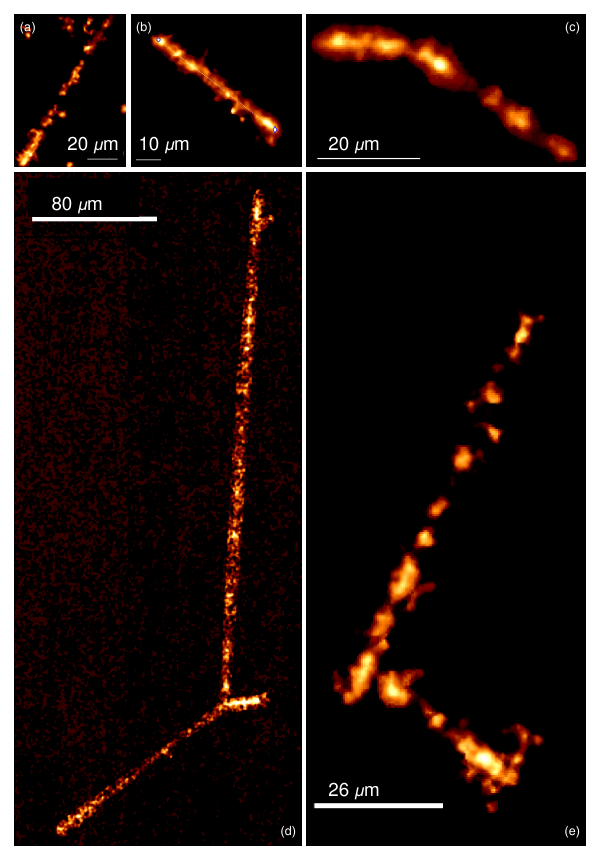}

     \caption{Cosmic-ray induced track candidates. Images in panels (a-c) and (e) are showing {\it processed z-stacks}, whereas panel (d) shows a pre-processed only z-stack.}
     \label{fig:cosmic}
 \end{figure}

 \section{Conclusion}

We quantified the dose response of LiF to fast-neutron induced recoils and gamma rays and found that it is dose-for-dose about 50 times less sensitive to gamma rays than neutron recoils. We also established the stability of the fluorescence over a period of several months. 

We demonstrated the three-dimensional imaging of particle-induced tracks of color centers in crystals using light-sheet fluorescence microscopy (LSFM). Specifically, we utilized the open-source mesoSPIM platform to image several crystals in response to various irradiation modes and obtained both qualitative and quantitative results from the acquired images, including estimates of the track sizes and topologies, which are indicative of the particle type. While we encountered some limitations, such as bleaching and camera noise, modifying a microscope with a low-noise camera will help acquiring images using the full capabilities of the mesoSPIM -- namely, its high acquisition speed and uniform axial resolution. Uniform illumination and a lower noise level will enable precise calorimetry and determination of $dE/dx$, aiding in particle identification.

We utilized open-source software and packages to perform data analysis. We predict a large improvement in correctly segmenting, classifying, and measuring tracks once the signal-to-noise ratio is increased and the axial resolution made uniform, where both can be achieved by a low-noise camera. The development of a fully integrated custom machine learning pipeline is underway to fully automate data reduction and event reconstruction. Scan times can be reduced to a few hours per cubic centimeter with hardware improvements. In combination, this will allow for the detection of single events in cubic centimeter-sized crystals.

The morphology, size, and number of the different event types, including those caused by thermal neutron capture on $^6$Li, conform to simulation results obtained with TRIM.  We also have demonstrated cosmic ray detection, and a selection of more exotic events is shown in Fig.~\ref{fig:cosmic}. Already, with the current performance, neutron detection is highly efficient, and neutron energy and direction reconstruction are currently being developed for applications in nuclear security~\cite{Cogswell:2021qlq} and arms control~\cite{Glaser2014}.
 
The fact that we can reliably detect triton tracks indicates that even low linear-energy transfer radiation that leaves sparse color center tracks is detectable even at the relatively high noise levels of the currently used camera. A low-noise camera will very likely enable \emph{single} color center detection capability.  Detection of reactor neutrino coherent elastic neutrino nucleus scattering (CEvNS) has recently been achieved by the CONUS+ experiment using active germanium detectors~\cite{Ackermann:2025obx}.  Single color center sensitivity will pave the way for passive reactor neutrino CEvNS detection. Underground deployment of a customized mesoSPIM microscope would allow us to establish the background levels for a direct dark matter search in the GeV and sub-GeV mass range with potentially competitive sensitivities to spin-dependent dark matter interactions~\cite{Cogswell:2021qlq}.

An important next step is the search for materials besides LiF for which this technique is effective -- especially if naturally abundant materials like quartz or diamond could be shown to work, paleo-detection of dark matter~\cite{Baum:2021jak,Baum:2023cct} could become a potential application as well.

\acknowledgements

This work was supported by the U.S. Department of Energy National Nuclear Security Administration Office of Defense Nuclear Nonproliferation R\&D through the Consortium for Monitoring, Technology, and Verification under award number DE-NA0003920 and by the National Science Foundation Growing Convergence Research award 2428507. The work of Gabriela R. Araujo was supported by the UZH Postdoc Grant No.K-72312-14-01. The mesoSPIM project and imaging platform were supported by the University Research Priority Program (URPP) “Adaptive Brain Circuits in Development and Learning (AdaBD)” of the University of Zurich. This work was supported by the Virginia Tech College of Science Lay Nam Chang Dean's Discovery Fund. This work was prepared by LLNL under Contract DE-AC52-07NA27344 and supported by the LLNL-LDRD Program under Project No. 23-FS-028. LLNL-JRNL-872914. This work  is supported by the European Union's Horizon Europe research and innovation programme under the Marie Skłodowska-Curie Postdoctoral Fellowship Programme, SMASH co-funded under the grant agreement No. 101081355. The SMASH project is co-funded by the Republic of Slovenia and the European Union from the European Regional Development Fund.

\bibliography{biblio,merged}

\begin{thebibliography}{30}%
\makeatletter
\providecommand \@ifxundefined [1]{%
 \@ifx{#1\undefined}
}%
\providecommand \@ifnum [1]{%
 \ifnum #1\expandafter \@firstoftwo
 \else \expandafter \@secondoftwo
 \fi
}%
\providecommand \@ifx [1]{%
 \ifx #1\expandafter \@firstoftwo
 \else \expandafter \@secondoftwo
 \fi
}%
\providecommand \natexlab [1]{#1}%
\providecommand \enquote  [1]{``#1''}%
\providecommand \bibnamefont  [1]{#1}%
\providecommand \bibfnamefont [1]{#1}%
\providecommand \citenamefont [1]{#1}%
\providecommand \href@noop [0]{\@secondoftwo}%
\providecommand \href [0]{\begingroup \@sanitize@url \@href}%
\providecommand \@href[1]{\@@startlink{#1}\@@href}%
\providecommand \@@href[1]{\endgroup#1\@@endlink}%
\providecommand \@sanitize@url [0]{\catcode `\\12\catcode `\$12\catcode `\&12\catcode `\#12\catcode `\^12\catcode `\_12\catcode `\%12\relax}%
\providecommand \@@startlink[1]{}%
\providecommand \@@endlink[0]{}%
\providecommand \url  [0]{\begingroup\@sanitize@url \@url }%
\providecommand \@url [1]{\endgroup\@href {#1}{\urlprefix }}%
\providecommand \urlprefix  [0]{URL }%
\providecommand \Eprint [0]{\href }%
\providecommand \doibase [0]{http://dx.doi.org/}%
\providecommand \selectlanguage [0]{\@gobble}%
\providecommand \bibinfo  [0]{\@secondoftwo}%
\providecommand \bibfield  [0]{\@secondoftwo}%
\providecommand \translation [1]{[#1]}%
\providecommand \BibitemOpen [0]{}%
\providecommand \bibitemStop [0]{}%
\providecommand \bibitemNoStop [0]{.\EOS\space}%
\providecommand \EOS [0]{\spacefactor3000\relax}%
\providecommand \BibitemShut  [1]{\csname bibitem#1\endcsname}%
\let\auto@bib@innerbib\@empty
\bibitem [{\citenamefont {Gruber}\ \emph {et~al.}(1997)\citenamefont {Gruber} \emph {et~al.}}]{NVconfocal}%
  \BibitemOpen
  \bibfield  {author} {\bibinfo {author} {\bibfnamefont {A.}~\bibnamefont {Gruber}} \emph {et~al.},\ }\bibfield  {title} {\enquote {\bibinfo {title} {Scanning confocal optical microscopy and magnetic resonance on single defect centers},}\ }\href {\doibase 10.1126/science.276.5321.2012} {\bibfield  {journal} {\bibinfo  {journal} {Science}\ }\textbf {\bibinfo {volume} {276}},\ \bibinfo {pages} {2012--2014} (\bibinfo {year} {1997})},\ \Eprint {http://arxiv.org/abs/https://www.science.org/doi/pdf/10.1126/science.276.5321.2012} {https://www.science.org/doi/pdf/10.1126/science.276.5321.2012} \BibitemShut {NoStop}%
\bibitem [{\citenamefont {Cogswell}\ \emph {et~al.}(2021)\citenamefont {Cogswell}, \citenamefont {Goel},\ and\ \citenamefont {Huber}}]{Cogswell:2021qlq}%
  \BibitemOpen
  \bibfield  {author} {\bibinfo {author} {\bibfnamefont {Bernadette~K.}\ \bibnamefont {Cogswell}}, \bibinfo {author} {\bibfnamefont {Apurva}\ \bibnamefont {Goel}}, \ and\ \bibinfo {author} {\bibfnamefont {Patrick}\ \bibnamefont {Huber}},\ }\bibfield  {title} {\enquote {\bibinfo {title} {Passive low-energy nuclear-recoil detection with color centers},}\ }\href {\doibase 10.1103/PhysRevApplied.16.064060} {\bibfield  {journal} {\bibinfo  {journal} {Phys. Rev. Applied}\ }\textbf {\bibinfo {volume} {16}},\ \bibinfo {pages} {064060} (\bibinfo {year} {2021})},\ \Eprint {http://arxiv.org/abs/2104.13926} {arXiv:2104.13926 [physics.ins-det]} \BibitemShut {NoStop}%
\bibitem [{\citenamefont {Akselrod}\ \emph {et~al.}(2006)\citenamefont {Akselrod}, \citenamefont {Akselrod}, \citenamefont {Benton},\ and\ \citenamefont {Yasuda}}]{Akselrod:2006a}%
  \BibitemOpen
  \bibfield  {author} {\bibinfo {author} {\bibfnamefont {G.M.}\ \bibnamefont {Akselrod}}, \bibinfo {author} {\bibfnamefont {M.S.}\ \bibnamefont {Akselrod}}, \bibinfo {author} {\bibfnamefont {E.R.}\ \bibnamefont {Benton}}, \ and\ \bibinfo {author} {\bibfnamefont {N.}~\bibnamefont {Yasuda}},\ }\bibfield  {title} {\enquote {\bibinfo {title} {A novel \protect{Al2O3} fluorescent nuclear track detector for heavy charged particles and neutrons},}\ }\href@noop {} {\bibfield  {journal} {\bibinfo  {journal} {Nucl. Instr. and Meth. B}\ }\textbf {\bibinfo {volume} {247}},\ \bibinfo {pages} {295--306} (\bibinfo {year} {2006})}\BibitemShut {NoStop}%
\bibitem [{\citenamefont {Akselrod}\ and\ \citenamefont {Sykora}(2011)}]{Akselrod:2011}%
  \BibitemOpen
  \bibfield  {author} {\bibinfo {author} {\bibfnamefont {M.S.}\ \bibnamefont {Akselrod}}\ and\ \bibinfo {author} {\bibfnamefont {G.J.}\ \bibnamefont {Sykora}},\ }\bibfield  {title} {\enquote {\bibinfo {title} {Fluorescent nuclear track detector technology - a new way to do passive solid state dosimetry},}\ }\href@noop {} {\bibfield  {journal} {\bibinfo  {journal} {Radiat. Measur.}\ }\textbf {\bibinfo {volume} {46}},\ \bibinfo {pages} {1671--1679} (\bibinfo {year} {2011})}\BibitemShut {NoStop}%
\bibitem [{\citenamefont {Akselrod}\ and\ \citenamefont {Kouwenbergfif}(2018)}]{Akselrod:2018}%
  \BibitemOpen
  \bibfield  {author} {\bibinfo {author} {\bibfnamefont {M.}~\bibnamefont {Akselrod}}\ and\ \bibinfo {author} {\bibfnamefont {J.}~\bibnamefont {Kouwenbergfif}},\ }\bibfield  {title} {\enquote {\bibinfo {title} {Fluorescent nuclear track detectors – review of past, present and future of the technology},}\ }\href@noop {} {\bibfield  {journal} {\bibinfo  {journal} {Radiat. Meas.}\ }\textbf {\bibinfo {volume} {117}},\ \bibinfo {pages} {35--51} (\bibinfo {year} {2018})}\BibitemShut {NoStop}%
\bibitem [{\citenamefont {Akselrod}\ \emph {et~al.}(2020)\citenamefont {Akselrod}, \citenamefont {V.},\ and\ \citenamefont {Harrison}}]{Akselrod:2020}%
  \BibitemOpen
  \bibfield  {author} {\bibinfo {author} {\bibfnamefont {M.}~\bibnamefont {Akselrod}}, \bibinfo {author} {\bibfnamefont {Fomenko}\ \bibnamefont {V.}}, \ and\ \bibinfo {author} {\bibfnamefont {J.}~\bibnamefont {Harrison}},\ }\bibfield  {title} {\enquote {\bibinfo {title} {Latest advances in \protect{FNTD} technology and instrumentation},}\ }\href@noop {} {\bibfield  {journal} {\bibinfo  {journal} {Radiat. Meas.}\ }\textbf {\bibinfo {volume} {133}},\ \bibinfo {pages} {106302} (\bibinfo {year} {2020})}\BibitemShut {NoStop}%
\bibitem [{\citenamefont {Bilski}\ and\ \citenamefont {Marczewska}(2017)}]{Bilski:2017}%
  \BibitemOpen
  \bibfield  {author} {\bibinfo {author} {\bibfnamefont {P.}~\bibnamefont {Bilski}}\ and\ \bibinfo {author} {\bibfnamefont {B.}~\bibnamefont {Marczewska}},\ }\bibfield  {title} {\enquote {\bibinfo {title} {Fluorescent detection of single tracks of alpha particles using lithium fluoride crystals},}\ }\href@noop {} {\bibfield  {journal} {\bibinfo  {journal} {Nucl. Instr. and Meth. B}\ }\textbf {\bibinfo {volume} {392}},\ \bibinfo {pages} {41--45} (\bibinfo {year} {2017})}\BibitemShut {NoStop}%
\bibitem [{\citenamefont {Bilski}\ \emph {et~al.}(2018)\citenamefont {Bilski}, \citenamefont {Marczewska}, \citenamefont {Kłosowski}, \citenamefont {M.},\ and\ \citenamefont {Naruszewicz}}]{Bilski:2018}%
  \BibitemOpen
  \bibfield  {author} {\bibinfo {author} {\bibfnamefont {P.}~\bibnamefont {Bilski}}, \bibinfo {author} {\bibfnamefont {B.}~\bibnamefont {Marczewska}}, \bibinfo {author} {\bibnamefont {Kłosowski}}, \bibinfo {author} {\bibfnamefont {W.}~\bibnamefont {M.}, \bibfnamefont {Gieszczyk}}, \ and\ \bibinfo {author} {\bibfnamefont {M.}~\bibnamefont {Naruszewicz}},\ }\bibfield  {title} {\enquote {\bibinfo {title} {Detection of neutrons with \protect{LiF} fluorescent nuclear track detectors},}\ }\href@noop {} {\bibfield  {journal} {\bibinfo  {journal} {Radiat. Meas.}\ }\textbf {\bibinfo {volume} {116}},\ \bibinfo {pages} {35--39} (\bibinfo {year} {2018})}\BibitemShut {NoStop}%
\bibitem [{\citenamefont {Bilski}\ \emph {et~al.}(2019)\citenamefont {Bilski}, \citenamefont {Marczewska}, \citenamefont {Gieszczyk}, \citenamefont {Kłosowski}, \citenamefont {Naruszewicz}, \citenamefont {Sankowska},\ and\ \citenamefont {Kodaira}}]{Bilksi:2019a}%
  \BibitemOpen
  \bibfield  {author} {\bibinfo {author} {\bibfnamefont {P.}~\bibnamefont {Bilski}}, \bibinfo {author} {\bibfnamefont {B.}~\bibnamefont {Marczewska}}, \bibinfo {author} {\bibfnamefont {W.}~\bibnamefont {Gieszczyk}}, \bibinfo {author} {\bibfnamefont {M.}~\bibnamefont {Kłosowski}}, \bibinfo {author} {\bibfnamefont {M.}~\bibnamefont {Naruszewicz}}, \bibinfo {author} {\bibfnamefont {M.}~\bibnamefont {Sankowska}}, \ and\ \bibinfo {author} {\bibfnamefont {S.}~\bibnamefont {Kodaira}},\ }\bibfield  {title} {\enquote {\bibinfo {title} {Fluorescent imaging of heavy charged particle tracks with \protect{LiF} single crystals},}\ }\href@noop {} {\bibfield  {journal} {\bibinfo  {journal} {Journal of Luminescence}\ }\textbf {\bibinfo {volume} {213}},\ \bibinfo {pages} {82--87} (\bibinfo {year} {2019})}\BibitemShut {NoStop}%
\bibitem [{\citenamefont {P.}\ \emph {et~al.}(2019)\citenamefont {P.}, \citenamefont {B.}, \citenamefont {W.}, \citenamefont {Kłosowski}, \citenamefont {Naruszewicz}, \citenamefont {M.},\ and\ \citenamefont {Y.}}]{Bilski:2019b}%
  \BibitemOpen
  \bibfield  {author} {\bibinfo {author} {\bibfnamefont {Bilksi}\ \bibnamefont {P.}}, \bibinfo {author} {\bibfnamefont {Marczewska}\ \bibnamefont {B.}}, \bibinfo {author} {\bibfnamefont {Gieszczyk}\ \bibnamefont {W.}}, \bibinfo {author} {\bibfnamefont {M.}~\bibnamefont {Kłosowski}}, \bibinfo {author} {\bibnamefont {Naruszewicz}}, \bibinfo {author} {\bibfnamefont {Zhydachevskyy}\ \bibnamefont {M.}}, \ and\ \bibinfo {author} {\bibfnamefont {S.}~\bibnamefont {Y.}, \bibfnamefont {Kodaira}},\ }\bibfield  {title} {\enquote {\bibinfo {title} {Luminescent properties of \protect{LiF} crystals for fluorescent imaging of nuclear particles tracks},}\ }\href@noop {} {\bibfield  {journal} {\bibinfo  {journal} {Optical Materials}\ }\textbf {\bibinfo {volume} {90}},\ \bibinfo {pages} {1--6} (\bibinfo {year} {2019})}\BibitemShut {NoStop}%
\bibitem [{\citenamefont {Bilski}\ \emph {et~al.}(2024)\citenamefont {Bilski}, \citenamefont {Marczewska}, \citenamefont {Sankowska}, \citenamefont {Kilian}, \citenamefont {Swako\'n}, \citenamefont {Siketi\'c},\ and\ \citenamefont {Olko}}]{Bilski:2024ghu}%
  \BibitemOpen
  \bibfield  {author} {\bibinfo {author} {\bibfnamefont {P.}~\bibnamefont {Bilski}}, \bibinfo {author} {\bibfnamefont {B.}~\bibnamefont {Marczewska}}, \bibinfo {author} {\bibfnamefont {M.}~\bibnamefont {Sankowska}}, \bibinfo {author} {\bibfnamefont {A.}~\bibnamefont {Kilian}}, \bibinfo {author} {\bibfnamefont {J.}~\bibnamefont {Swako\'n}}, \bibinfo {author} {\bibfnamefont {Z.}~\bibnamefont {Siketi\'c}}, \ and\ \bibinfo {author} {\bibfnamefont {P.}~\bibnamefont {Olko}},\ }\bibfield  {title} {\enquote {\bibinfo {title} {{Detection of proton tracks with \protect{LiF} fluorescent nuclear track detectors}},}\ }\href {\doibase 10.1016/j.radmeas.2024.107083} {\bibfield  {journal} {\bibinfo  {journal} {Radiat. Meas.}\ }\textbf {\bibinfo {volume} {173}},\ \bibinfo {pages} {107083} (\bibinfo {year} {2024})},\ \Eprint {http://arxiv.org/abs/2403.04320} {arXiv:2403.04320 [physics.ins-det]} \BibitemShut {NoStop}%
\bibitem [{\citenamefont {Perez}\ \emph {et~al.}(2024)\citenamefont {Perez}, \citenamefont {Walkup}, \citenamefont {Chapman}, \citenamefont {Bhaumik}, \citenamefont {Khodaparast}, \citenamefont {Magill}, \citenamefont {Huber},\ and\ \citenamefont {Ivanov}}]{Perez:2024hly}%
  \BibitemOpen
  \bibfield  {author} {\bibinfo {author} {\bibfnamefont {Mariano~Guerrero}\ \bibnamefont {Perez}}, \bibinfo {author} {\bibfnamefont {Keegan}\ \bibnamefont {Walkup}}, \bibinfo {author} {\bibfnamefont {Jordan}\ \bibnamefont {Chapman}}, \bibinfo {author} {\bibfnamefont {Pranshu}\ \bibnamefont {Bhaumik}}, \bibinfo {author} {\bibfnamefont {Giti~A.}\ \bibnamefont {Khodaparast}}, \bibinfo {author} {\bibfnamefont {Brenden~A.}\ \bibnamefont {Magill}}, \bibinfo {author} {\bibfnamefont {Patrick}\ \bibnamefont {Huber}}, \ and\ \bibinfo {author} {\bibfnamefont {Vsevolod}\ \bibnamefont {Ivanov}},\ }\bibfield  {title} {\enquote {\bibinfo {title} {{First-principles Spin and Optical Properties of Vacancy Clusters in Lithium Fluoride}},}\ }\href@noop {} {\  (\bibinfo {year} {2024})},\ \Eprint {http://arxiv.org/abs/2412.21060} {arXiv:2412.21060 [cond-mat.mtrl-sci]} \BibitemShut {NoStop}%
\bibitem [{\citenamefont {Vladimirov}\ \emph {et~al.}(2024)\citenamefont {Vladimirov}, \citenamefont {Voigt}, \citenamefont {Naert}, \citenamefont {Araujo}, \citenamefont {Cai}, \citenamefont {Reuss}, \citenamefont {Zhao}, \citenamefont {Schmid}, \citenamefont {Hildebrand}, \citenamefont {Schaettin}, \citenamefont {Groos}, \citenamefont {Mateos}, \citenamefont {Bethge}, \citenamefont {Yamamoto}, \citenamefont {Aerne}, \citenamefont {Roebroeck}, \citenamefont {Ert{\"u}rk}, \citenamefont {Aguzzi}, \citenamefont {Ziegler}, \citenamefont {Stoeckli}, \citenamefont {Baudis}, \citenamefont {Lienkamp},\ and\ \citenamefont {Helmchen}}]{Vladimirov:2024}%
  \BibitemOpen
  \bibfield  {author} {\bibinfo {author} {\bibfnamefont {Nikita}\ \bibnamefont {Vladimirov}}, \bibinfo {author} {\bibfnamefont {Fabian~F.}\ \bibnamefont {Voigt}}, \bibinfo {author} {\bibfnamefont {Thomas}\ \bibnamefont {Naert}}, \bibinfo {author} {\bibfnamefont {Gabriela~R.}\ \bibnamefont {Araujo}}, \bibinfo {author} {\bibfnamefont {Ruiyao}\ \bibnamefont {Cai}}, \bibinfo {author} {\bibfnamefont {Anna~Maria}\ \bibnamefont {Reuss}}, \bibinfo {author} {\bibfnamefont {Shan}\ \bibnamefont {Zhao}}, \bibinfo {author} {\bibfnamefont {Patricia}\ \bibnamefont {Schmid}}, \bibinfo {author} {\bibfnamefont {Sven}\ \bibnamefont {Hildebrand}}, \bibinfo {author} {\bibfnamefont {Martina}\ \bibnamefont {Schaettin}}, \bibinfo {author} {\bibfnamefont {Dominik}\ \bibnamefont {Groos}}, \bibinfo {author} {\bibfnamefont {Jos{\'e}~Mar{\'\i}a}\ \bibnamefont {Mateos}}, \bibinfo {author} {\bibfnamefont {Philipp}\ \bibnamefont {Bethge}}, \bibinfo {author} {\bibfnamefont {Taiyo}\ \bibnamefont {Yamamoto}}, \bibinfo {author} {\bibfnamefont
  {Valentino}\ \bibnamefont {Aerne}}, \bibinfo {author} {\bibfnamefont {Alard}\ \bibnamefont {Roebroeck}}, \bibinfo {author} {\bibfnamefont {Ali}\ \bibnamefont {Ert{\"u}rk}}, \bibinfo {author} {\bibfnamefont {Adriano}\ \bibnamefont {Aguzzi}}, \bibinfo {author} {\bibfnamefont {Urs}\ \bibnamefont {Ziegler}}, \bibinfo {author} {\bibfnamefont {Esther}\ \bibnamefont {Stoeckli}}, \bibinfo {author} {\bibfnamefont {Laura}\ \bibnamefont {Baudis}}, \bibinfo {author} {\bibfnamefont {Soeren~S.}\ \bibnamefont {Lienkamp}}, \ and\ \bibinfo {author} {\bibfnamefont {Fritjof}\ \bibnamefont {Helmchen}},\ }\bibfield  {title} {\enquote {\bibinfo {title} {Benchtop mesospim: a next-generation open-source light-sheet microscope for cleared samples},}\ }\href {\doibase 10.1038/s41467-024-46770-2} {\bibfield  {journal} {\bibinfo  {journal} {Nat. Commun.}\ }\textbf {\bibinfo {volume} {15}},\ \bibinfo {pages} {2679} (\bibinfo {year} {2024})}\BibitemShut {NoStop}%
\bibitem [{\citenamefont {Agostinelli}\ \emph {et~al.}(2003)\citenamefont {Agostinelli} \emph {et~al.}}]{GEANT4}%
  \BibitemOpen
  \bibfield  {author} {\bibinfo {author} {\bibfnamefont {S.}~\bibnamefont {Agostinelli}} \emph {et~al.} (\bibinfo {collaboration} {GEANT4}),\ }\bibfield  {title} {\enquote {\bibinfo {title} {{GEANT4 - A Simulation Toolkit}},}\ }\href {\doibase 10.1016/S0168-9002(03)01368-8} {\bibfield  {journal} {\bibinfo  {journal} {Nucl. Instrum. Meth. A}\ }\textbf {\bibinfo {volume} {506}},\ \bibinfo {pages} {250--303} (\bibinfo {year} {2003})}\BibitemShut {NoStop}%
\bibitem [{\citenamefont {Ziegler}\ \emph {et~al.}(2010)\citenamefont {Ziegler}, \citenamefont {Ziegler},\ and\ \citenamefont {Biersack}}]{TRIM}%
  \BibitemOpen
  \bibfield  {author} {\bibinfo {author} {\bibfnamefont {James}\ \bibnamefont {Ziegler}}, \bibinfo {author} {\bibfnamefont {M.D.}\ \bibnamefont {Ziegler}}, \ and\ \bibinfo {author} {\bibfnamefont {J.}~\bibnamefont {Biersack}},\ }\bibfield  {title} {\enquote {\bibinfo {title} {The stopping and range of ions in mater},}\ }\href {\doibase 10.1016/j.nimb.2010.02.091} {\bibfield  {journal} {\bibinfo  {journal} {NIM B}\ }\textbf {\bibinfo {volume} {268}},\ \bibinfo {pages} {1818--1823} (\bibinfo {year} {2010})}\BibitemShut {NoStop}%
\bibitem [{amb(2001)}]{ambe}%
  \BibitemOpen
  \href@noop {} {\emph {\bibinfo {title} {Compendium of neutron spectra and detector responses for radiation protection purposes : supplement to technical reports series no. 318.}}},\ \bibinfo {type} {Tech. Rep.}\ (\bibinfo  {institution} {International Atomic Energy Agency},\ \bibinfo {year} {2001})\ \bibinfo {note} {\protect{STI/DOC/010/403}}\BibitemShut {NoStop}%
\bibitem [{\citenamefont {Sears}(1992)}]{Sears01011992}%
  \BibitemOpen
  \bibfield  {author} {\bibinfo {author} {\bibfnamefont {Varley~F.}\ \bibnamefont {Sears}},\ }\bibfield  {title} {\enquote {\bibinfo {title} {Neutron scattering lengths and cross sections},}\ }\href@noop {} {\bibfield  {journal} {\bibinfo  {journal} {Neutron News}\ }\textbf {\bibinfo {volume} {3}},\ \bibinfo {pages} {26--37} (\bibinfo {year} {1992})}\BibitemShut {NoStop}%
\bibitem [{\citenamefont {Ziegler}(1996)}]{Ziegler:1996}%
  \BibitemOpen
  \bibfield  {author} {\bibinfo {author} {\bibfnamefont {J.~F.}\ \bibnamefont {Ziegler}},\ }\bibfield  {title} {\enquote {\bibinfo {title} {Terrestrial cosmic rays},}\ }\href {\doibase 10.1147/rd.401.0019} {\bibfield  {journal} {\bibinfo  {journal} {IBM Journal of Research and Development}\ }\textbf {\bibinfo {volume} {40}},\ \bibinfo {pages} {19--39} (\bibinfo {year} {1996})}\BibitemShut {NoStop}%
\bibitem [{\citenamefont {Ju}\ \emph {et~al.}(2021)\citenamefont {Ju}, \citenamefont {Lin}, \citenamefont {Shen}, \citenamefont {Wu},\ and\ \citenamefont {Wu}}]{doi:10.1080/23746149.2020.1858721}%
  \BibitemOpen
  \bibfield  {author} {\bibinfo {author} {\bibfnamefont {Zhiping}\ \bibnamefont {Ju}}, \bibinfo {author} {\bibfnamefont {Junjie}\ \bibnamefont {Lin}}, \bibinfo {author} {\bibfnamefont {Si}~\bibnamefont {Shen}}, \bibinfo {author} {\bibfnamefont {Botao}\ \bibnamefont {Wu}}, \ and\ \bibinfo {author} {\bibfnamefont {E}~\bibnamefont {Wu}},\ }\bibfield  {title} {\enquote {\bibinfo {title} {Preparations and applications of single color centers in diamond},}\ }\href {\doibase 10.1080/23746149.2020.1858721} {\bibfield  {journal} {\bibinfo  {journal} {Advances in Physics: X}\ }\textbf {\bibinfo {volume} {6}},\ \bibinfo {pages} {1858721} (\bibinfo {year} {2021})}\BibitemShut {NoStop}%
\bibitem [{\citenamefont {Sow}\ \emph {et~al.}(2020)\citenamefont {Sow}, \citenamefont {Steuer}, \citenamefont {Adekanye}, \citenamefont {Ginés}, \citenamefont {Mandal}, \citenamefont {Gilboa}, \citenamefont {Williams}, \citenamefont {Smith},\ and\ \citenamefont {Kapanidis}}]{D0NR05931E}%
  \BibitemOpen
  \bibfield  {author} {\bibinfo {author} {\bibfnamefont {Maabur}\ \bibnamefont {Sow}}, \bibinfo {author} {\bibfnamefont {Horst}\ \bibnamefont {Steuer}}, \bibinfo {author} {\bibfnamefont {Sanmi}\ \bibnamefont {Adekanye}}, \bibinfo {author} {\bibfnamefont {Laia}\ \bibnamefont {Ginés}}, \bibinfo {author} {\bibfnamefont {Soumen}\ \bibnamefont {Mandal}}, \bibinfo {author} {\bibfnamefont {Barak}\ \bibnamefont {Gilboa}}, \bibinfo {author} {\bibfnamefont {Oliver~A.}\ \bibnamefont {Williams}}, \bibinfo {author} {\bibfnamefont {Jason~M.}\ \bibnamefont {Smith}}, \ and\ \bibinfo {author} {\bibfnamefont {Achillefs~N.}\ \bibnamefont {Kapanidis}},\ }\bibfield  {title} {\enquote {\bibinfo {title} {High-throughput nitrogen-vacancy center imaging for nanodiamond photophysical characterization and ph nanosensing},}\ }\href {\doibase 10.1039/D0NR05931E} {\bibfield  {journal} {\bibinfo  {journal} {Nanoscale}\ }\textbf {\bibinfo {volume} {12}},\ \bibinfo {pages} {21821--21831} (\bibinfo {year} {2020})}\BibitemShut {NoStop}%
\bibitem [{web()}]{webmesospim}%
  \BibitemOpen
  \href {http://mesospim.org} {\enquote {\bibinfo {title} {{The mesoSPIM initiative website}},}\ }\bibinfo {note} {Retrieved in August 2023}\BibitemShut {NoStop}%
\bibitem [{\citenamefont {Voigt}\ \emph {et~al.}(2019)\citenamefont {Voigt}, \citenamefont {Kirschenbaum}, \citenamefont {Platonova}, \citenamefont {Pag{\`e}s}, \citenamefont {Campbell}, \citenamefont {Kastli}, \citenamefont {Schaettin}, \citenamefont {Egolf}, \citenamefont {Van Der~Bourg}, \citenamefont {Bethge} \emph {et~al.}}]{Voigt2019}%
  \BibitemOpen
  \bibfield  {author} {\bibinfo {author} {\bibfnamefont {Fabian~F}\ \bibnamefont {Voigt}}, \bibinfo {author} {\bibfnamefont {Daniel}\ \bibnamefont {Kirschenbaum}}, \bibinfo {author} {\bibfnamefont {Evgenia}\ \bibnamefont {Platonova}}, \bibinfo {author} {\bibfnamefont {St{\'e}phane}\ \bibnamefont {Pag{\`e}s}}, \bibinfo {author} {\bibfnamefont {Robert~AA}\ \bibnamefont {Campbell}}, \bibinfo {author} {\bibfnamefont {Rahel}\ \bibnamefont {Kastli}}, \bibinfo {author} {\bibfnamefont {Martina}\ \bibnamefont {Schaettin}}, \bibinfo {author} {\bibfnamefont {Ladan}\ \bibnamefont {Egolf}}, \bibinfo {author} {\bibfnamefont {Alexander}\ \bibnamefont {Van Der~Bourg}}, \bibinfo {author} {\bibfnamefont {Philipp}\ \bibnamefont {Bethge}},  \emph {et~al.},\ }\bibfield  {title} {\enquote {\bibinfo {title} {The mesospim initiative: open-source light-sheet microscopes for imaging cleared tissue},}\ }\href@noop {} {\bibfield  {journal} {\bibinfo  {journal} {Nature Methods}\ }\textbf {\bibinfo {volume} {16}},\ \bibinfo {pages}
  {1105--1108} (\bibinfo {year} {2019})}\BibitemShut {NoStop}%
\bibitem [{\citenamefont {Berg}\ \emph {et~al.}(2019)\citenamefont {Berg}, \citenamefont {Kutra}, \citenamefont {Kroeger}, \citenamefont {Straehle}, \citenamefont {Kausler}, \citenamefont {Haubold}, \citenamefont {Schiegg}, \citenamefont {Ales}, \citenamefont {Beier}, \citenamefont {Rudy}, \citenamefont {Eren}, \citenamefont {Cervantes}, \citenamefont {Xu}, \citenamefont {Beuttenmueller}, \citenamefont {Wolny}, \citenamefont {Zhang}, \citenamefont {Koethe}, \citenamefont {Hamprecht},\ and\ \citenamefont {Kreshuk}}]{berg2019}%
  \BibitemOpen
  \bibfield  {author} {\bibinfo {author} {\bibfnamefont {Stuart}\ \bibnamefont {Berg}}, \bibinfo {author} {\bibfnamefont {Dominik}\ \bibnamefont {Kutra}}, \bibinfo {author} {\bibfnamefont {Thorben}\ \bibnamefont {Kroeger}}, \bibinfo {author} {\bibfnamefont {Christoph~N.}\ \bibnamefont {Straehle}}, \bibinfo {author} {\bibfnamefont {Bernhard~X.}\ \bibnamefont {Kausler}}, \bibinfo {author} {\bibfnamefont {Carsten}\ \bibnamefont {Haubold}}, \bibinfo {author} {\bibfnamefont {Martin}\ \bibnamefont {Schiegg}}, \bibinfo {author} {\bibfnamefont {Janez}\ \bibnamefont {Ales}}, \bibinfo {author} {\bibfnamefont {Thorsten}\ \bibnamefont {Beier}}, \bibinfo {author} {\bibfnamefont {Markus}\ \bibnamefont {Rudy}}, \bibinfo {author} {\bibfnamefont {Kemal}\ \bibnamefont {Eren}}, \bibinfo {author} {\bibfnamefont {Jaime~I.}\ \bibnamefont {Cervantes}}, \bibinfo {author} {\bibfnamefont {Buote}\ \bibnamefont {Xu}}, \bibinfo {author} {\bibfnamefont {Fynn}\ \bibnamefont {Beuttenmueller}}, \bibinfo {author} {\bibfnamefont {Adrian}\
  \bibnamefont {Wolny}}, \bibinfo {author} {\bibfnamefont {Chong}\ \bibnamefont {Zhang}}, \bibinfo {author} {\bibfnamefont {Ullrich}\ \bibnamefont {Koethe}}, \bibinfo {author} {\bibfnamefont {Fred~A.}\ \bibnamefont {Hamprecht}}, \ and\ \bibinfo {author} {\bibfnamefont {Anna}\ \bibnamefont {Kreshuk}},\ }\bibfield  {title} {\enquote {\bibinfo {title} {ilastik: interactive machine learning for (bio)image analysis},}\ }\href@noop {} {\bibfield  {journal} {\bibinfo  {journal} {Nature Methods}\ }\textbf {\bibinfo {volume} {16}},\ \bibinfo {pages} {1226–1232} (\bibinfo {year} {2019})}\BibitemShut {NoStop}%
\bibitem [{\citenamefont {van~der Walt}\ \emph {et~al.}(2014)\citenamefont {van~der Walt}, \citenamefont {{S}ch\"onberger}, \citenamefont {{Nunez-Iglesias}}, \citenamefont {{B}oulogne}, \citenamefont {{W}arner}, \citenamefont {{Y}ager}, \citenamefont {{G}ouillart}, \citenamefont {{Y}u},\ and\ \citenamefont {the scikit-image contributors}}]{scikit-image}%
  \BibitemOpen
  \bibfield  {author} {\bibinfo {author} {\bibfnamefont {{S}t\'efan}\ \bibnamefont {van~der Walt}}, \bibinfo {author} {\bibfnamefont {{J}ohannes~{L}.}\ \bibnamefont {{S}ch\"onberger}}, \bibinfo {author} {\bibfnamefont {{J}uan}\ \bibnamefont {{Nunez-Iglesias}}}, \bibinfo {author} {\bibfnamefont {{F}ran\c{c}ois}\ \bibnamefont {{B}oulogne}}, \bibinfo {author} {\bibfnamefont {{J}oshua~{D}.}\ \bibnamefont {{W}arner}}, \bibinfo {author} {\bibfnamefont {{N}eil}\ \bibnamefont {{Y}ager}}, \bibinfo {author} {\bibfnamefont {{E}mmanuelle}\ \bibnamefont {{G}ouillart}}, \bibinfo {author} {\bibfnamefont {{T}ony}\ \bibnamefont {{Y}u}}, \ and\ \bibinfo {author} {\bibnamefont {the scikit-image contributors}},\ }\bibfield  {title} {\enquote {\bibinfo {title} {scikit-image: image processing in {P}ython},}\ }\href@noop {} {\bibfield  {journal} {\bibinfo  {journal} {PeerJ}\ }\textbf {\bibinfo {volume} {2}},\ \bibinfo {pages} {e453} (\bibinfo {year} {2014})}\BibitemShut {NoStop}%
\bibitem [{\citenamefont {Imhof}\ \emph {et~al.}(1959)\citenamefont {Imhof}, \citenamefont {Johnson}, \citenamefont {Vaughn},\ and\ \citenamefont {Walt}}]{PhysRev.114.1037}%
  \BibitemOpen
  \bibfield  {author} {\bibinfo {author} {\bibfnamefont {W.~L.}\ \bibnamefont {Imhof}}, \bibinfo {author} {\bibfnamefont {R.~G.}\ \bibnamefont {Johnson}}, \bibinfo {author} {\bibfnamefont {F.~J.}\ \bibnamefont {Vaughn}}, \ and\ \bibinfo {author} {\bibfnamefont {M.}~\bibnamefont {Walt}},\ }\bibfield  {title} {\enquote {\bibinfo {title} {Cross sections for the \protect{${\mathrm{Li}}^{7}(n, \ensuremath{\gamma}){\mathrm{Li}}^{8}$} reaction},}\ }\href {\doibase 10.1103/PhysRev.114.1037} {\bibfield  {journal} {\bibinfo  {journal} {Phys. Rev.}\ }\textbf {\bibinfo {volume} {114}},\ \bibinfo {pages} {1037--1039} (\bibinfo {year} {1959})}\BibitemShut {NoStop}%
\bibitem [{\citenamefont {Brown}\ \emph {et~al.}(1963)\citenamefont {Brown}, \citenamefont {James}, \citenamefont {Perkin},\ and\ \citenamefont {Barry}}]{BROWN1963137}%
  \BibitemOpen
  \bibfield  {author} {\bibinfo {author} {\bibfnamefont {F.}~\bibnamefont {Brown}}, \bibinfo {author} {\bibfnamefont {R.H.}\ \bibnamefont {James}}, \bibinfo {author} {\bibfnamefont {J.L.}\ \bibnamefont {Perkin}}, \ and\ \bibinfo {author} {\bibfnamefont {J.}~\bibnamefont {Barry}},\ }\bibfield  {title} {\enquote {\bibinfo {title} {The cross section of the \protect{7Li(n, t)} reaction for neutron energies between 3.5 and 15 \protect{MeV}},}\ }\href {\doibase https://doi.org/10.1016/0368-3230(63)90118-6} {\bibfield  {journal} {\bibinfo  {journal} {Journal of Nuclear Energy. Parts A/B. Reactor Science and Technology}\ }\textbf {\bibinfo {volume} {17}},\ \bibinfo {pages} {137--141} (\bibinfo {year} {1963})}\BibitemShut {NoStop}%
\bibitem [{\citenamefont {Glaser}\ \emph {et~al.}(2014)\citenamefont {Glaser}, \citenamefont {Barak},\ and\ \citenamefont {Goldston}}]{Glaser2014}%
  \BibitemOpen
  \bibfield  {author} {\bibinfo {author} {\bibfnamefont {A.}~\bibnamefont {Glaser}}, \bibinfo {author} {\bibfnamefont {B.}~\bibnamefont {Barak}}, \ and\ \bibinfo {author} {\bibfnamefont {R.}~\bibnamefont {Goldston}},\ }\bibfield  {title} {\enquote {\bibinfo {title} {{A zero-knowledge protocol for nuclear warhead verification}},}\ }\href@noop {} {\bibfield  {journal} {\bibinfo  {journal} {Nature}\ }\textbf {\bibinfo {volume} {510}},\ \bibinfo {pages} {497--502} (\bibinfo {year} {2014})}\BibitemShut {NoStop}%
\bibitem [{\citenamefont {Ackermann}\ \emph {et~al.}(2025)\citenamefont {Ackermann} \emph {et~al.}}]{Ackermann:2025obx}%
  \BibitemOpen
  \bibfield  {author} {\bibinfo {author} {\bibfnamefont {N.}~\bibnamefont {Ackermann}} \emph {et~al.},\ }\bibfield  {title} {\enquote {\bibinfo {title} {{First observation of reactor antineutrinos by coherent scattering}},}\ }\href@noop {} {\  (\bibinfo {year} {2025})},\ \Eprint {http://arxiv.org/abs/2501.05206} {arXiv:2501.05206 [hep-ex]} \BibitemShut {NoStop}%
\bibitem [{\citenamefont {Baum}\ \emph {et~al.}(2021)\citenamefont {Baum}, \citenamefont {Edwards}, \citenamefont {Freese},\ and\ \citenamefont {Stengel}}]{Baum:2021jak}%
  \BibitemOpen
  \bibfield  {author} {\bibinfo {author} {\bibfnamefont {Sebastian}\ \bibnamefont {Baum}}, \bibinfo {author} {\bibfnamefont {Thomas D.~P.}\ \bibnamefont {Edwards}}, \bibinfo {author} {\bibfnamefont {Katherine}\ \bibnamefont {Freese}}, \ and\ \bibinfo {author} {\bibfnamefont {Patrick}\ \bibnamefont {Stengel}},\ }\bibfield  {title} {\enquote {\bibinfo {title} {{New Projections for Dark Matter Searches with Paleo-Detectors}},}\ }\href {\doibase 10.3390/instruments5020021} {\bibfield  {journal} {\bibinfo  {journal} {Instruments}\ }\textbf {\bibinfo {volume} {5}},\ \bibinfo {pages} {21} (\bibinfo {year} {2021})},\ \Eprint {http://arxiv.org/abs/2106.06559} {arXiv:2106.06559 [astro-ph.CO]} \BibitemShut {NoStop}%
\bibitem [{\citenamefont {Baum}\ \emph {et~al.}(2023)\citenamefont {Baum} \emph {et~al.}}]{Baum:2023cct}%
  \BibitemOpen
  \bibfield  {author} {\bibinfo {author} {\bibfnamefont {Sebastian}\ \bibnamefont {Baum}} \emph {et~al.},\ }\bibfield  {title} {\enquote {\bibinfo {title} {{Mineral detection of neutrinos and dark matter. A whitepaper}},}\ }\href {\doibase 10.1016/j.dark.2023.101245} {\bibfield  {journal} {\bibinfo  {journal} {Phys. Dark Univ.}\ }\textbf {\bibinfo {volume} {41}},\ \bibinfo {pages} {101245} (\bibinfo {year} {2023})},\ \Eprint {http://arxiv.org/abs/2301.07118} {arXiv:2301.07118 [astro-ph.IM]} \BibitemShut {NoStop}%
\end{thebibliography}%

\newpage
\section*{Appendix}

This appendix contains the x-y-projection of all 943 objects identified in sample 0505. Note, that the categorization of events is based on their 3-dimensional brightness distribution, but owing to the difficulty of plotting this information in full only the x-y-projections are shown. Dark colors correspond to high brightness.

\begin{figure}[h]
    \centering
    \includegraphics[width=\linewidth]{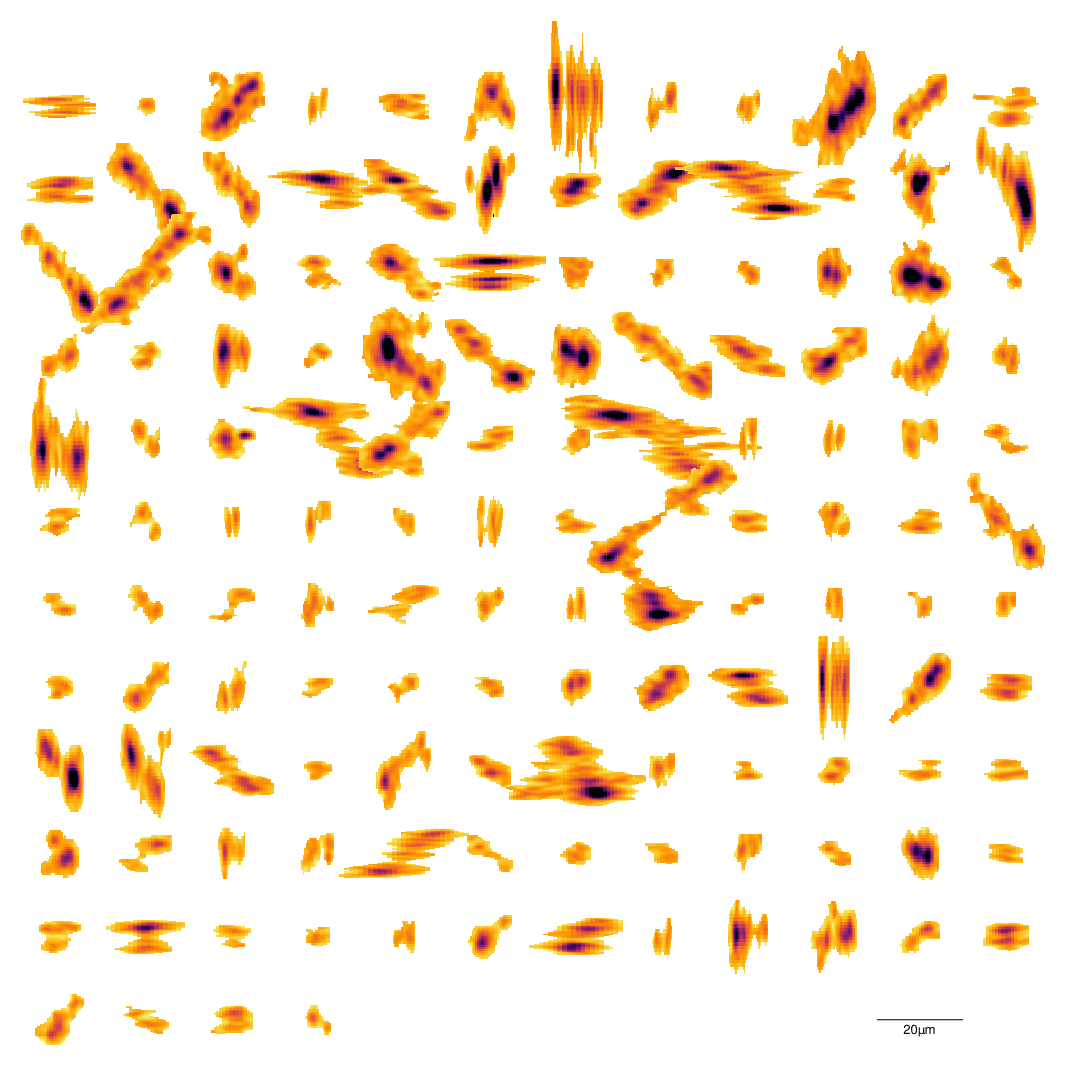}
    \caption{Extended, thermal neutron capture event candidates from sample 0505.}
    
\end{figure}

\begin{figure}[p]
    \centering
    \includegraphics[width=\linewidth]{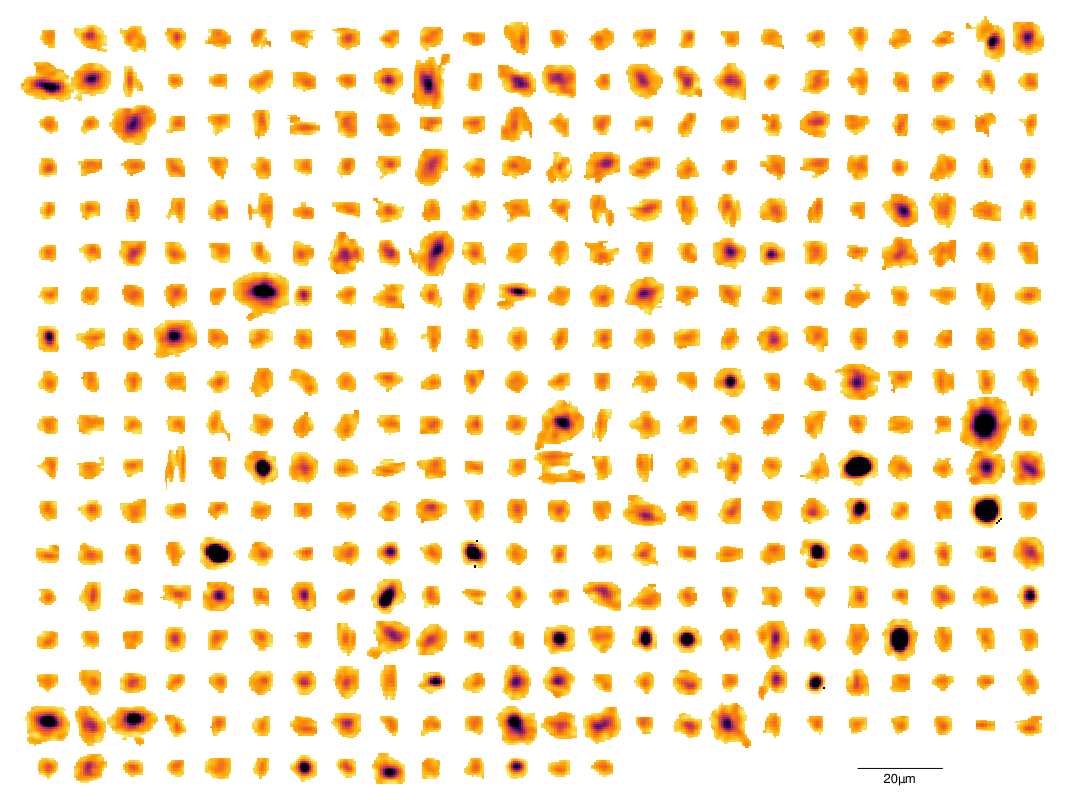}
    \caption{Compact, fast neutron recoil event candidates from sample 0505.}
    
\end{figure}

\begin{figure}[p]
    \centering
    \includegraphics[width=\linewidth]{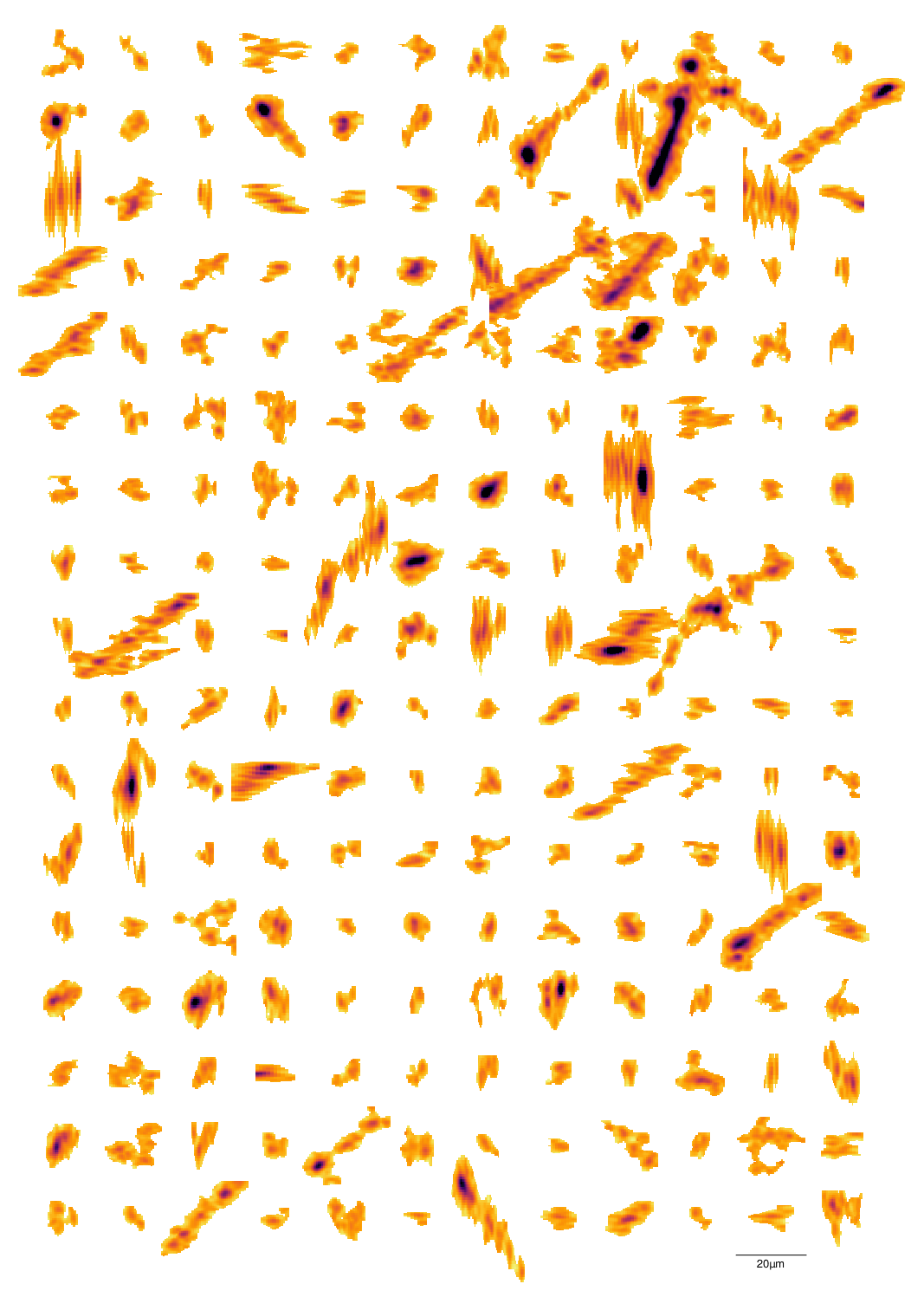}
    \caption{Unidentified event candidates from sample 0505.}
    
\end{figure}

\begin{figure}[p]
    \centering
    \includegraphics[width=\linewidth]{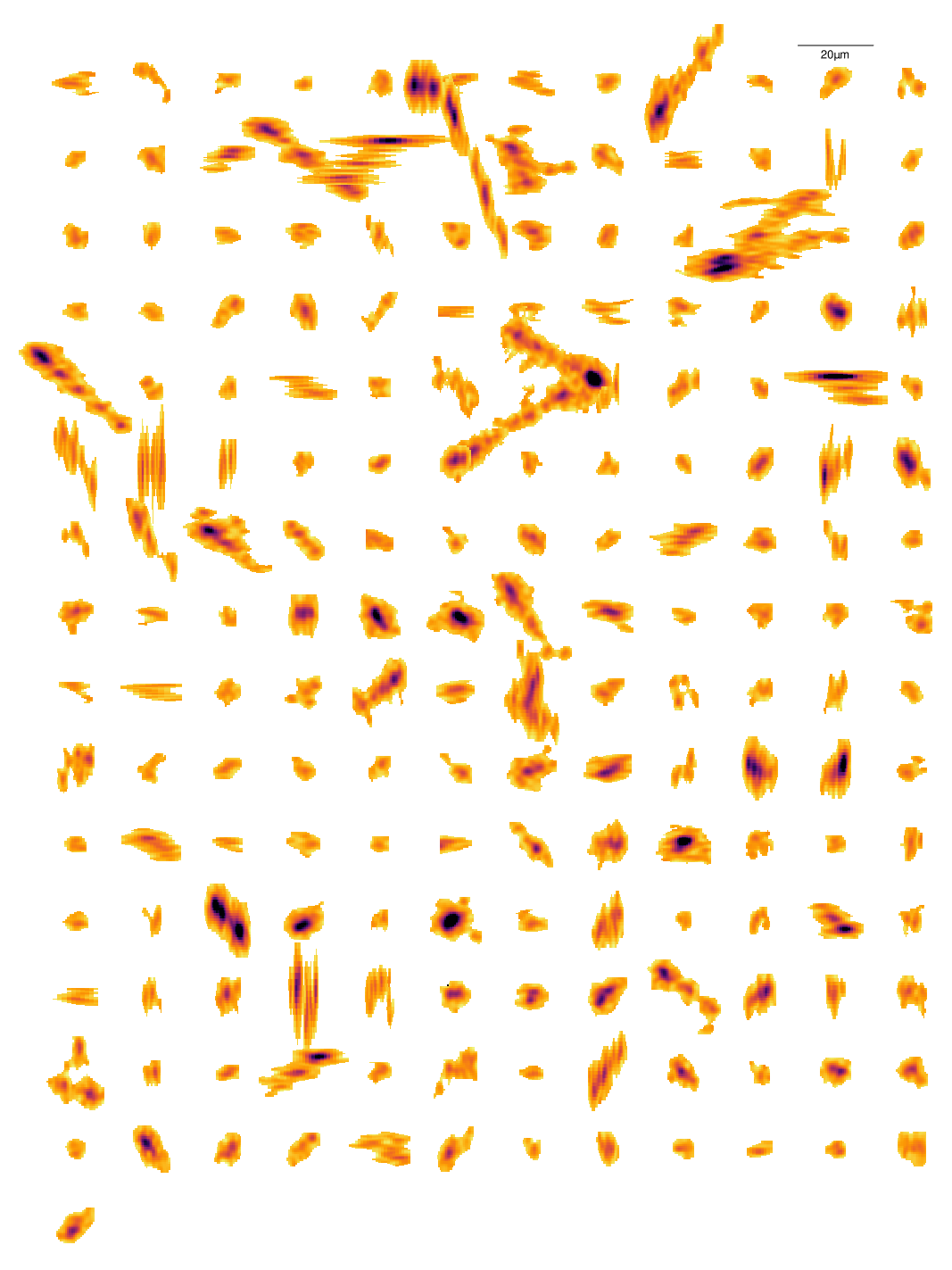}
    \caption{Unidentified event candidates from sample 0505.}
\end{figure}

\end{document}